\documentclass[11pt,twoside]{article}

\newcommand{\eps}{\varepsilon}
\begin{document}
\thispagestyle{empty}

\leftline{\copyright~ 1998 International Press}
\leftline{Adv. Theor. Math. Phys. {\bf 2} (1998) 1075-1103}  
\vspace{0.4in}
\begin{center}\renewcommand{\thefootnote}{\fnsymbol{footnote}}
{\huge \bf On The Dynamics of Einstein's}

{\huge \bf Equations in The Ashtekar}

{\huge \bf Formulation\footnote{Work supported by CONICOR, CONICET and Se.CyT, UNC} }
\vspace{0.1in}

{\bf $^{a}$Mirta S. Iriondo\footnote{Partially supported by AIT,C\'ordoba} , Enzo O. $^{a}$Leguizam\'on\footnote{Fellow of Se.CyT-UNC.}, Oscar A. $^{a,b}$Reula\footnote{Member of CONICET.}}
\linebreak
\renewcommand{\thefootnote}{\arabic{footnote}}
\renewcommand{\thefootnote}{}
\footnotetext{\small e-print archive: {\texttt http://xxx.lanl.gov/abs/gr-qc/9804019}}

$^{a}$FaMAF, Medina Allende y Haya de la Torre,\\
Ciudad Universitaria, \\
5000 C\'ordoba\\
Argentina\\ 
$\:$  \\
$^{b}$Albert-Einstein-Institut\\
Max-Planck-Institut f\"ur Gravitationsphysik\\
Schlaatzweg 1\\
14473 Potsdam\\
Germany\\

\renewcommand{\thefootnote}{\arabic{footnote}}
\setcounter{footnote}{0}
%\title{On the dynamics of Einstein's equations in the Ashtekar
%formulation.~\footnote{Work supported
%             by CONICOR, CONICET and Se.CyT, UNC}}

%\author{{\sc Mirta S. Iriondo}\thanks{Partially supported by AIT,
%C\'ordoba,
%    Argentina. }, {\sc Enzo O. Leguizam\'on} \thanks{Fellow of
%    Se.CyT-UNC.} \\
%{\small FaMAF, Medina Allende y Haya de la Torre,}\\
%  {\small Ciudad Universitaria, 5000 C\'ordoba, Argentina}\\
%   {\sc Oscar A. Reula}
%  \thanks{Member of CONICET.}\\
%  {\small FaMAF, Medina Allende y Haya de la Torre,}\\
%  {\small Ciudad Universitaria, 5000 C\'ordoba, Argentina}\\
%  {\small and} \\
%  {\small Albert-Einstein-Institut} \\
%  {\small Max-Planck-Institut f\"ur Gravitationsphysik} \\
%  {\small Schlaatzweg 1, 14473 Potsdam, Germany} \\
%}
\vspace{0.1in} 
{\bf Abstract} \\
\vspace{0.1in}

%\maketitle \pacs{...}

%\vspace{-.4in}
%\begin{center}
%  {\bf Abstract}
%\end{center}
%\begin{abstract}

\parbox[c]{4.5 in}{\small {\hspace{0.2 in}} We study the dynamics of Einstein's equations in Ashtekar's
  variables from the point of view of the theory of hyperbolic systems
  of evolution equations. We extend previous results and show that by
  a suitable modification of the Hamiltonian vector flow outside the
  sub-manifold of real and constrained solutions, a symmetric
  hyperbolic system is obtained for any fixed choice of lapse-shift
  pair, without assuming the solution to be a priori real.  We notice
  that the evolution system is block diagonal in the pair
  $(\sigma^a,A_b)$, and provide explicit and very simple formulae for
  the eigenvector-eigenvalue pairs in terms of an orthonormal tetrad
  with one of its components pointing along the propagation direction.
  We also analyze the constraint equations and find that when viewed as
  functions of the extended phase space they form a
  symmetric hyperbolic system on their own. We also provide simple
  formulae for its eigenvectors-eigenvalues pairs.}
\end{center}
\newpage
\pagenumbering{arabic}
\setcounter{page}{1076}

\pagestyle{myheadings}
\markboth{\it ON THE DYNAMICS OF EINSTEIN'S EQUATIONS}{\it M. IRIONDO, E. LEGUIZAM\'ON, O. REULA }
\section{Introduction}
\label{sec:int}

A substantial effort  has been undertaken in the last decade to
understand what a canonical quantization of space-time would mean in
terms of self-dual variables and the corresponding loop representation
it implies.  Most of the effort has been geared to the study of the
constraint equations
and their algebra, while little has been done about  the evolution
equations, in the understanding that, since the Hamiltonian of the
theory is a linear combination of the constraints, the information
contained in their algebra is all what is needed.

Nevertheless we believe that the study of the evolution equations as
such should be
of relevance for a better understanding of the constraint algebra, for
it
might assert that it has very important properties which might otherwise
be very difficult to recognize, namely that of giving rise to a well
posed initial value formulation, that is, a theory in which one could
predict the future based on knowledge gathered at an initial Cauchy
surface.

It is not clear a priori   that a classical well posed initial value
formulation is essential for quantization. However all physically interesting quantum theories we know of correspond to  well posed classical
systems, and this fact is implicitly used in most approximation schemes.

The well posedness of the initial value problem for the classical theory
in the usual tensorial variables
was asserted in the fifties by Choquet-Bruhat\cite{cho} using a particular
reduction of the equations in the harmonic gauge. Since then, a number of other formulations of the theory as symmetric hyperbolic systems
have appeared (see \cite{cho,cho-rugg,helmut,fri-reu,bonna-masso,cho-york,putten} and for an updated review with a more complete
reference list see \cite{livrev} ), mainly
aiming to gain some advantages in dealing with certain specific
problems, such as asymptotic studies, global existence, the Newtonian
limit and numerical simulations of fully general relativistic
configurations. 

It is not clear that the study of the dynamics of Einstein's equations
in Ashtekar's variables as a symmetric hyperbolic system  would have an
important impact on  applications such as those   mentioned above, but its intrinsic beauty,
simplicity, and economy   can not be in vane.

In dealing with first order quasi-linear systems of evolution
equations, --as Einstein's equations naturally are in Ashtekar's
variables-- a sufficient condition for well posedness is symmetric
hyperbolicity, that is, if the system can be written in the form:
\begin{equation}
M^0(u) u_t := M^a(u) \nabla_a u + B(u)u
\end{equation}
where $u$ is a ``vector'' and $M^0$, $M^a$ and $B$ are matrices which
have smooth dependence on $u$ and, in addition   $M^0$ and $M^a$ are
symmetric and  $M^0$ is positive definite~\footnote{For a
short introduction into
the topic see \cite{livrev} and references therein.}. 
This condition is the one of more immediate
use, in particular for numerical simulations, for it enables one to estimate
the growth of the solution in terms of conventional norms; therefore it is
always worth checking whether or not it holds. If a system does not
have this property it does not mean that its associated initial value
formulation is not well posed, for the necessary condition for well
posedness is a
weaker condition, namely strong hyperbolicity~\footnote{The name strong
comes
because it is stronger than the primitive concept of hyperbolicity,
namely that the eigenvalues of the associated symbol be purely
imaginary, but it is a weaker condition than symmetric hyperbolicity.},  
that is, that the eigenvalues of the principal symbol,
$iM^a(u)k_a$ be purely imaginary and that it has a complete set of
eigenvectors, all of them smooth in their dependence on $u$ and
$k_a$. It turns out that the Ashtekar's system can be extended outside
the constraint and reality sub-manifold of the phase space in such a way that both
hyperbolicity
conditions hold.

In determining whether a formulation of general relativity is well
posed or not there usually arises the problem of constraints, since we
can not locally separate the fields into free and into dependent ones with
respect to the constraints. We have to think on the evolution
equations not only as giving the dynamics at the constraint
sub-manifolds, but also as evolution equations for the complete fields,
that is, as equations also valid in a neighborhood of the constraint
sub-manifold.  
But this makes those equations non-uniquely determined; all dynamics
which
at the constraint sub-manifold give the same evolution vector field are
 valid, and therefore equivalent. Thus, if we modify the equations, by adding to
them
terms proportional to the constraints, we obtain equivalent systems of
equations.  Thus, in order to show well posedness one only has to
search for one of these equivalent systems of equations in which the
above conditions hold. 

In the Ashtekar formulation of general
relativity a further complication arises because besides the
constraints there are reality conditions. Along evolution the
soldering forms are in general not anti-hermitian, even if they started
that way at 
the initial surface, so without assuming very stringent evolution
gauge conditions
or imposing elliptic gauges, one can only require the metric reality
conditions along evolution and since these conditions are non-linear in
terms of 
the basic dynamical variables, it becomes
necessary to deal with them as if they were a new set of
constraints.

In a recent paper \cite{Iriondo}, the authors have answered the
question of \newpage \noindent hyperbolicity by finding a representative within the
equivalent class of evolution equations and showing that it is a
symmetric hyperbolic system.  In that paper it was assumed that the
soldering form was anti-hermitian,  restricting unnecessarily  the
set of gauge conditions (choices of lapse-shift pair and of the
${\bf SU}(2)$ rotation gauge) for which symmetric hyperbolicity holds.
That
restriction can be easily lifted to allow for all soldering forms
to be\newline  ${\bf SL}(2,{\bf C})$ rotated to an anti-hermitian one,
that is, for
all soldering forms satisfying the metric reality condition. We do this in the
next section, (\S \ref{sec:sym}).

Unfortunately this lifting is not enough to grant the well posedness
of the system, because a small departure from the reality sub-manifold
would imply an immediate and strong instability for certain
perturbations.  Thus, the usual methods of constructing the solution
via a contraction map of successive approximations can not be used,
nor can one use these equations for numerical simulations.  To
overcome this problem a further modification to the evolution
equations is needed, this time outside the reality condition
sub-manifold, which makes the system symmetric-hyperbolic in a whole
neighborhood of that sub-manifold. We study this modification in  \S \ref{sec:FR}.
 In order to distinguish it from the
prior modifications above mentioned, we shall call it {\sl flow
  regularization.}

We then study, in \S \ref{sec:CD}, the diagonalization
problem for the principal symbol of the system, that is, the issue of
strong hyperbolicity and of the propagation speeds of perturbations.
In particular we display remarkably simple expressions for the
different eigenvalues-eigenvectors of the system. We conclude that the
system can not be extended outside the reality condition sub-manifold
in order to make it strongly hyperbolic there, keeping at the same
time unchanged the simple dependence of the characteristic
eigenvectors on the solution.  On the other hand one can modify the
equation into a strongly hyperbolic pseudo-differential system, the
advantage here being that the characteristics of the system remain
very simple.

In \S \ref{ev-const-real} we study the issue of the constraint and reality
conditions propagation, showing that the system stays on those
sub-manifolds if initially so.  We show  in particular that the
constraints propagate also via a symmetric hyperbolic system. The
diagonalization of this system is also remarkably simple.

%%%%%%%%%%

%%%%%%%%%%

\section{Symmetric Hyperbolicity}
\label{sec:sym}
%%%%%%%%%%%%%%%%%%%%%%%%%%%%%%%%%%%

\subsection{Earlier Results}
\label{sec:paper1}

In a recent paper \cite{Iriondo}, the authors have  shown that the Einstein's equations in Ashtekar's variables constitute   a
symmetric hyperbolic system.  In that paper it was assumed that the
soldering form was anti-hermitian~\footnote{Since we are using
intensely the matrix algebra of the soldering forms we refer to them as
anti-hermitian
in the matrix sense, in the abstract sense, as defined in \cite{ash}
they are actually hermitian.},  restricting unnecessarily  the
set of gauge conditions  for which symmetric hyperbolicity holds.

In order to obtain a symmetric hyperbolic
 system for any given lapse-shift pair, we suitably extended
the field equations outside the constraint sub-manifold. 
Since this extension only involves the addition of terms proportional
to the constraints, we know that the evolution vector field is tangent to
the  constraint sub-manifold, and so the physically relevant 
evolution, that is, the dynamics inside this sub-manifold,
remains unchanged. 

 For completeness,  we repeat here part of the calculations made in  \cite{Iriondo}, showing in addition that the     above extension is    unique.

It is known (\cite{ash}, see appendix \ref{sec:ash} for a brief overview) that, using the Hamiltonian formulation to determine the dynamics  of the pair $(\tilde\sigma^a, A_b)$, Einstein's equations  can be written as
\begin{eqnarray}
\label{eq:complexeinstein}
{\cal L}_t\tilde\sigma^b&=&{\cal L}_{\vec
N}\tilde\sigma^b-\frac{i}{\sqrt{2}}{\cal D}_a(\;\hbox
{${}_{{}_{{}_{\widetilde{}}}}$\kern-.5em\it
N}[\tilde\sigma^{a},\tilde\sigma^{b}]) ,\nonumber\\
{\cal L}_t A_b&=&{\cal L}_{\vec N}A_b+\frac{i}{\sqrt{2}}\;\;\hbox
{${}_{{}_{{}_{\widetilde{}}}}$\kern-.5em\it
N}[\tilde\sigma^a,F_{ab}],\nonumber\\
C(\tilde\sigma,A)&:=&\mbox{tr}(\tilde\sigma^a\tilde\sigma^bF_{ab})=0,\\
C_a(\tilde\sigma,A)&:=&\mbox{tr}(\tilde\sigma^bF_{ab})=0,\nonumber
\\
\tilde C_A{}^B(\tilde\sigma,A)&:=&{\cal D}_a
\tilde\sigma^a{}_A{}^B=0,\nonumber
\end{eqnarray}

Since the   evolution equations  have a block diagonal
principal part, that is, in the time evolution for $\tilde{\sigma}^a$
($A_a$) there only appear
space derivatives of $\tilde{\sigma}^a$ (respectively $A_a$), the desired 
modification  is of the following form:~\footnote{One can always add to these 
equations {\sl given} ${\bf SU}(2,{\bf C})$ gauge terms in the usual way without affecting
hyperbolicity, for if they do not depend on $\tilde{\sigma}^a$, nor on $A_a$, they do not
enter the principal part.}
 
\begin{eqnarray}
\label{eq:evolution1S}
{\cal L}_t\tilde\sigma^b&=&{\cal L}_{\vec
N}\tilde\sigma^b-\frac{i}{\sqrt{2}}{\cal D}_a(\;\hbox
{${}_{{}_{{}_{\widetilde{}}}}$\kern-.5em\it
N}[\tilde\sigma^{a},\tilde\sigma^{b}]) +\alpha [\tilde
C,\tilde\sigma^b],\\
\label{eq:evolution1A}
{\cal L}_t A_b&=&{\cal L}_{\vec N}A_b+\frac{i}{\sqrt{2}}\;\;\hbox
{${}_{{}_{{}_{\widetilde{}}}}$\kern-.5em\it N}[\tilde\sigma^a,F_{ab}]
+\beta \;\tilde\sigma_b C+\gamma \;\epsilon_b{}^{dc}\tilde\sigma_c C_d.
\end{eqnarray}
where $\alpha, \beta$ and $\gamma$ are complex functions to be
determined so  that the system becomes symmetric-hyperbolic, meaning 
that the principal symbol becomes anti-hermitian with respect to the 
canonical inner product
\begin{eqnarray*}
  \langle u_2, u_1\rangle &\equiv
&\langle(\tilde\sigma_2,A^2),(\tilde\sigma_1,A^1)\rangle\\
  &\equiv & \mbox{tr}(\tilde\sigma^{\dag}_2{}^a
  \tilde\sigma_1^b)\;q_{ab}+ \mbox{tr}(A^{2\dag}{}_a
  A^{1}{}_b)\;q^{ab},
\end{eqnarray*}
and considering the soldering forms, $\tilde{\sigma}^a$  
anti-hermitian.

Recall that the principal  symbol of a quasi-linear evolution equation system
$$
\dot u^l=B^l{}_j{}^a(u)\nabla_a u^j + M^l(u)
$$
is given by $P(u,ik_a)=i\;B^l{}_j{}^a(u) k_a$. In our case $u$ denotes $u=(\sigma^a,A_b)$. Thus we need to prove that
$$
P_{12}+P^{\dag}{}_{12}\equiv \langle u_2,Pu_1\rangle + \langle u_2,P^{\dag}  u_1\rangle=0.
$$
Using the fact that any complex matrix can be written as $u =
-\frac1{\sigma^2}\mbox{tr}(u\tilde{\sigma}_e)\tilde{\sigma}^e$ and that 
$[\tilde{\sigma}^a,\tilde{\sigma}^b] =
\sigma \sqrt{2} \eps^{ab}{}_c\tilde{\sigma}^c$, 
the principal symbol can be written as

\begin{eqnarray}
\label{eq:op(sigma)1}
P_\sigma(\tilde\sigma,ik_a)\tilde\sigma_1{}^b&=&-i\frac{k_a
N^a}{\sigma^2}\makebox{tr}(\tilde\sigma_1{}^b\tilde\sigma_e)\tilde\sigma^e
- \frac{k_a}{\sigma  }\;\;\hbox{${}_{{}_{{}_{\widetilde{}}}}$\kern-.5em\rm N}
\mbox{tr}(\tilde{\sigma}_1{}^b\tilde{\sigma}^e)\eps^{a}{}_{ec}\tilde{\sigma}^c\nonumber\\
&&-\frac{i\sqrt{2}}{\sigma}\left(\alpha-\frac{i}{\sqrt{2}}\;\hbox
{${}_{{}_{{}_{\widetilde{}}}}$\kern-.5em\rm N}\right)k_a
\mbox{tr}(\tilde\sigma_1{}^a\tilde\sigma_e)\eps^{eb}{}_{d}\tilde{
\sigma}^d,
\end{eqnarray}

\begin{eqnarray}
\label{eq:op(A)1}
P_A(\tilde\sigma,ik_c)\;A^1{}_b&=&-i\frac{k_a
N^a}{\sigma^2}\mbox{tr}(A_{1b}\tilde\sigma_e)\tilde\sigma^e +
\frac{2}{\sigma}\;\hbox {${}_{{}_{{}_{\widetilde{}}}}$\kern-.5em\rm N}
k_{[a} 
\mbox{tr}(A^1_{b]}\tilde{\sigma}_e)\eps^{ae}{}_c\tilde{\sigma}^c\nonumber\\
&&+i\;\beta \sqrt{2}\sigma \eps^{cde} k_c
\tilde\sigma_b\;\mbox{tr}(\tilde\sigma_e\;A^1_d)+i2\;\gamma
\eps_b{}^{dc} \tilde\sigma_c\;\mbox{tr}(\tilde\sigma^a\;
k_{[d}A^1_{a]}).
\end{eqnarray}
A very simple calculus shows, using the   inner product above and the
anti-hermiticity of the background soldering form, that
\begin{eqnarray}
&&P_{\sigma 12} + P^{\dag}{}_{\sigma 12}=-i\frac{k_a
N^a}{\sigma^2}\makebox{tr}(\tilde\sigma_1{}^b\tilde\sigma_e)\makebox{tr}(\tilde{\sigma}^{\dag}_{2b}\tilde\sigma^e)+i\frac{k_a
N^a}{\sigma^2}\makebox{tr}(\tilde\sigma_2{}^{\dag
b}\tilde\sigma_e)\makebox{tr}(\tilde\sigma_{1b}\tilde\sigma^e)\nonumber\\
&&\qquad\qquad-\frac{1}{\sigma}\;\hbox
{${}_{{}_{{}_{\widetilde{}}}}$\kern-.5em\rm
N}\;k_a\;\mbox{tr}(\tilde\sigma_1{}^b\tilde\sigma_e)\eps^{ae}{}_c\;\makebox{tr}(\tilde{\sigma}^{\dag}_{2b}\tilde\sigma^c)-\frac{1}{\sigma}\;\hbox
{${}_{{}_{{}_{\widetilde{}}}}$\kern-.5em\rm
N}\;k_a\;\mbox{tr}(\tilde\sigma_{1b}\tilde\sigma^e)\eps^{ac}{}_e\;\nonumber\\
&&\qquad\qquad\times\makebox{tr}(\tilde{\sigma}_2{}^{\dag
b}\tilde\sigma_c)%\qquad\qquad\qquad
-\frac{i\sqrt{2}}{\sigma}(\alpha-\frac{i}{\sqrt{2}}\;\hbox
{${}_{{}_{{}_{\widetilde{}}}}$\kern-.5em\rm
N})\;k_a\;\mbox{tr}(\tilde\sigma_1{}^a\tilde\sigma_e)\eps^{eb}{}_d\;\makebox{tr}(\tilde{\sigma}^{\dag}_{2b}\tilde\sigma^d)\nonumber\\
&&\qquad\qquad+\frac{i\sqrt{2}}{\sigma}(\bar{\alpha}+\frac{i}{\sqrt{2}}\;\hbox
{${}_{{}_{{}_{\widetilde{}}}}$\kern-.5em\rm
N})\;k_a\;\mbox{tr}(\tilde\sigma_{1b}\tilde\sigma^d)\eps^{eb}{}_d\;\makebox{tr}(\tilde{\sigma}_2{}^{\dag
a}\tilde\sigma_e).
\end{eqnarray}
The sum of the first and the second term vanishes because of the symmetry
of the background metric 
used in the inner product, the sum of the third and the fourth term
vanishes because of the 
anti-symmetry of $\eps^{abc}$, finally the last two terms should vanish
in order to get the 
desired result, but it is impossible for the sum to vanish because there
are different contractions 
on the indices and   $\tilde\sigma_1$ is independent of
$\tilde\sigma_2$, thus, the only way 
for these terms to vanish is that each one of them   do so separately. 
This is obtained requiring
$$
\alpha=\frac{i}{\sqrt{2}}\;\hbox
{${}_{{}_{{}_{\widetilde{}}}}$\kern-.5em\rm N}.
$$
Next we calculate
\begin{eqnarray}
&&P_{A 12} + P^{\dag}{}_{A
12}=-i\frac{k_aN^a}{\sigma^2}\mbox{tr}(A^1{}_b\tilde\sigma^e)\mbox{tr}(A^{\dag
2}{}^{b}\tilde\sigma_e)+i\frac{k_aN^a}{\sigma^2}\mbox{tr}(A^{1b}\tilde\sigma^e)\nonumber \\\ 
&&\times \mbox{tr}(A^{\dag 2}{}_{b}\tilde\sigma_e)
+\;\hbox {${}_{{}_{{}_{\widetilde{}}}}$\kern-.5em\rm
N}\;\frac{k_a}{\sigma}\eps^{aem}\mbox{tr}(A^1{}_{b}\tilde\sigma_e)\mbox{tr}(A^{\dag
2}{}^{b}\tilde\sigma_m)+\;\hbox
{${}_{{}_{{}_{\widetilde{}}}}$\kern-.5em\rm
N}\;\frac{k_a}{\sigma}\eps^{aem}\mbox{tr}(A^1{}^{b}\tilde\sigma_m)\nonumber \\
&&\mbox{tr}(A^{\dag
2}{}_{b}\tilde\sigma_e)+i\beta\sqrt{2}\sigma\eps^{cde}\;\mbox{tr}(A^{\dag
2}{}^{b}\tilde\sigma_b)k_c\;\mbox{tr}(A^{1}{}_{d}\tilde\sigma_e)-i\bar{\gamma}k_d\eps^{dc}{}_{b}\;\mbox{tr}(A^{\dag
2}{}^{a}\tilde\sigma_a)\;\nonumber \\
&&k_c\mbox{tr}(A^{1}{}^{b}\tilde\sigma_c)+i\gamma\eps^{dc}{}_{b}k_d\;\mbox{tr}(A^{\dag
2}{}^{b}\tilde\sigma_c)\;\mbox{tr}(A^{1}{}^{a}\tilde\sigma_a)-i\bar{\beta}\sqrt{2}\sigma\eps^{cde}k_c\;\mbox{tr}(A^{\dag
2}{}_{d}\tilde\sigma_e)\;\nonumber \\ 
&&\mbox{tr}(A^{1}{}^{b}\tilde\sigma_b)-\;\hbox {${}_{{}_{{}_{\widetilde{}}}}$\kern-.5em\rm
N}\;\frac{k_b}{\sigma}\eps^{apn}\;\mbox{tr}(A^{\dag
2}{}^{b}\tilde\sigma_n)\;\mbox{tr}(A^{1}{}_{a}\tilde\sigma_p)-i\gamma\eps_a{}^{dp}k_b\;\mbox{tr}(A^{\dag
2}{}^{a}\tilde\sigma_p)\;\nonumber \\ 
&&\mbox{tr}(A^{1}{}_{d}\tilde\sigma^p)-\;\hbox {${}_{{}_{{}_{\widetilde{}}}}$\kern-.5em\rm
N}\;\frac{k_b}{\sigma}\eps^{apn}\;\mbox{tr}(A^{\dag
2}{}_{a}\tilde\sigma_p)\;\mbox{tr}(A^{1}{}^{b}\tilde\sigma_n)+i\bar{\gamma}\eps_b{}^{dc}k_a\;\mbox{tr}(A^{\dag
2}{}_{d}\tilde\sigma^a)\;\nonumber \\
&&\mbox{tr}(A^{1}{}^{b}\tilde\sigma_c).
\end{eqnarray}
The sum of the first and the second term vanishes because of the symmetry
of the metric, the third and fourth terms vanish because of the anti-symmetry
of $\eps^{abc}$. The next four terms give rise to
\begin{eqnarray}
&&i\;\mbox{tr}(A^{2\dag}{}^b\tilde\sigma_b)\;\mbox{tr}(A^{1}{}_d\tilde\sigma_e)\eps^{cde}k_c(\beta\sqrt{2}\sigma+\bar{\gamma})+i\;\mbox{tr}(A^{2\dag}{}_d\tilde\sigma_e)\;\mbox{tr}(A^{1}{}_a\tilde\sigma^a)\eps^{dce}k_c\nonumber\\
&&\times(\bar{\beta}\sqrt{2}\sigma+\gamma)\nonumber
\end{eqnarray}
and there are no other terms of the same kind, so they must vanish. The only way 
(since $A^1$ and $A^2$ are independent) this can happens is if
$$
\beta\sqrt{2}\sigma=-\bar{\gamma}.
$$
Finally the last four terms vanish if we  use
$$
2 A^{[a}\tilde\sigma^{b]}=\eps^{abe}\eps_{dme}A^d\tilde\sigma^m,
$$
and choose
$$
\gamma=-\frac{i}{\sigma}\;\hbox
{${}_{{}_{{}_{\widetilde{}}}}$\kern-.5em\rm N}, \qquad \mbox{ and
consequently}\qquad \beta=-\frac{i}{\sigma^2\sqrt{2}}\;\hbox
{${}_{{}_{{}_{\widetilde{}}}}$\kern-.5em\rm N}.
$$

%%%%%%%%%%%%%%%%%%%%%%%%%%%%%%%%%%%%%%%%%%%%%%%%%%%%%%%%%%%%%%%%%%%%%%%%%%%%%%%%%%%%%%%%%%

\subsection{Lifting the Anti-Hermiticity Condition}
\label{subsec:LA}
%%%%%%%%%%%%%%%%%%%%%%%%%%%%%%%%%%%%%%%%%%%%%%%%%%%%%%%%%%%%%%%%%%%%%%%%%%%%%%%%%%%%%%%%%%

In the above calculations it was assumed that $\tilde{\sigma}^a$ was anti-hermitian.
A simple modification of the scalar product into a ``rotated'' one,
allows to conclude that the
equations are still symmetric hyperbolic (w.r.t. the new scalar product)
even when the soldering forms are not anti-hermitian, as long as they
can be transformed into ones that are so by a ${\bf SL}(2,{\bf C})$
transformation, that is, as long as the metric they generate is real
and positive definite. 

Indeed, given $\tilde{\sigma}^a$ such that the metric it produces is real, 
there exist ${\bf SL}(2,{\bf C})$ transformations $L(\tilde{\sigma}^a)$ such
that $\hat{\sigma}^a := L^{-1}\tilde{\sigma}^a L$ is anti-hermitian.
(See Appendix \ref{sec:L} for a procedure to construct one.)
Using any one of these transformations we define the following scalar product:
\begin{eqnarray}
\label{eq:scalar1}
\langle u_2, u_1\rangle_{H(u)} &\equiv &\langle
H^{-1}(\tilde\sigma)(\tilde\sigma_2,A^2),H(\tilde\sigma)
(\tilde\sigma_1,A^1)\rangle\\
&\equiv & \mbox{tr}((L^{-1}\tilde\sigma_2^a L)^{\dag}
\;L^{-1}\tilde\sigma_1^b L)\;{q}_{ab}+  
\mbox{tr}((L^{-1}A^2_a L)^{\dag}\; L^{-1} A^{1}_b L)\;{q}^{ab},\nonumber \\
\end{eqnarray}
where the linear operator $H(u)$ is defined through $L$ as in the second step. 
Thus, when computing
$$
P_{12}{}_{H}+P^{\dag}{}_{12}{}_{H}\equiv \mbox{tr}((L^{-1}
u_2L)^{\dag}L^{-1}Pu_1L) + \mbox{tr}((L^{-1}Pu_2L)^{\dag}L^{-1}u_1L),
$$
we obtain terms with the structure,\footnote{We consider the case $N^a=0$, for the part
proportional to it is diagonal, hence symmetric, without any
conditions on $\tilde{\sigma}^a$.}
\begin{eqnarray}
&&\mbox{tr}((L^{-1}u_2L)^{\dag} L^{-1}(L\hat{\sigma}L^{-1})L \mbox{tr}(u_1
L \hat{\sigma}L^{-1})) = \nonumber\\
&&\mbox{tr}((L^{-1}u_2L)^{\dag} \hat{\sigma}) \mbox{tr}(L^{-1}u_1 L
\hat{\sigma}),\nonumber
\end{eqnarray}
that is, with the same structure as the one in the calculation in
\cite{Iriondo}, and above, if one substitutes $u_1$, and $u_2$ by
their rotated versions, $L^{-1}u_1L$, and $L^{-1}u_2L$, and
correspondingly $\tilde{\sigma}^c$ by $\hat{\sigma}^c$. Since
$\hat{\sigma}^c$ is by construction anti-hermitian,
symmetric-hyperbolicity follows for the extended system from the same
calculations as in the previous subsection.

As argued below, symmetric hyperbolicity in the reality condition
sub-manifold does not seems to be enough for the usual proof of well
posedness to work.  Thus, a further modification of the equations is
needed to obtain a set of equations which are symmetric hyperbolic in
a whole neighborhood of the reality conditions sub-manifold. We do
this in the next section.

%%%%%%%%%%%%%%%%%%%%%%%%%%%%%%%%%%%%%%%%%%%%%%%%%%%%%%%%%%%%%%%%%%%%%%%%%%%

\section{Flow Regularization}
\label{sec:FR}
%%%%%%%%%%%%%%%%%%%%%%%%%%%%%%%%%%%%%%%%%%%%%%%%%%%%%%%%%%%%%%%%%%%%%%%%%%%

Although, as we shall prove in section \ref{ev-const-real}, the evolution 
respects the reality conditions (and therefore solutions whose initial
data sets satisfy them, have a metric which stays real along
evolution), this property is not sufficient to conclude that the system is well posed.
Indeed, all schemes used to prove existence of solutions rely upon
sequences of intermediate equations and their corresponding solutions,
which are shown to form a contractive map, and hence to converge to the exact 
solution. These intermediate
equations are in general linear versions of the original equations,
but with the principal part evaluated at earlier approximated
solutions.  The problem is that this intermediate solutions do not
propagate correctly the reality conditions and so one has to solve
approximated systems in a neighborhood of the reality condition
sub-manifold, but there the system is not symmetric hyperbolic, and so
its eigenvalues have non-zero real part. Thus, the intermediate
equations unstable, and so if intermediate solutions exist at all their
norm can not be properly estimated.

This problem is overcome by further modifying the evolution equations
outside the reality condition in such a way to obtain a symmetric hyperbolic
system even outside of it.  Then standard contracting maps schemes for
proving existence of solutions can then be applied to obtain local well
posedness.

Here we present a detailed description of the required extension to
the flow in order to obtain a symmetric hyperbolicity in a whole neighborhood 
of the reality condition sub-manifold.

If the metric is real and positive definite, that is, if it lies in the 
reality condition sub-manifold, then  to prove symmetric-hyperbolicity one can simply use the fact   that there is an ${\bf SL}(2,{\bf C})$ 
transformation which makes $\sigma$ anti-hermitian.
Outside the reality sub-manifold no extension of the transformation 
would give an anti-hermitian soldering form, so the system there is 
not symmetric-hyperbolic without a further modification. 

The extra modification of the evolution equations consists first in
extending the ${\bf SL}(2,{\bf C})$ transformation outside the reality
condition sub-manifold, with properties which ensure that when 
the reality conditions are satisfied the resulting $\tilde{\sigma^a}$
would be anti-hermitian; we present in detail one way to achieve this in Appendix \ref{sec:L} (the procedure is not unique).  The second step is to change   everywhere in the principal symbol of the system
$\tilde{\sigma^a}$ by $L\hat{\sigma}^aL^{-1}:=
\frac12(L(L^{-1}\tilde{\sigma}^a L - (L^{-1}\tilde{\sigma}^a
L)^{\dagger})L^{-1})$.  At the reality sub-manifold we re-obtain
$\tilde{\sigma}^a$ but, outside of it, the equations differ.  Notice
also that $\hat{\sigma}^a$ is, by construction, anti-hermitian.  
Therefore we claim that the system is
everywhere symmetrizable, with symmetrizer $H(\tilde{\sigma}):=
(L^{-1})^{\dagger}L^{-1}$ to the left and $H^{-1}$ to the right
(notice that $H$ is symmetric and positive definite everywhere, since
$L \in {\bf SL}(2,{\bf C})$), that is, with respect to the following scalar product:

\begin{eqnarray*}
\langle u_2, u_1\rangle_{H(u)} &\equiv &\langle
H^{-1}(\tilde\sigma)(\tilde\sigma_2,A^2),H(\tilde\sigma)
(\tilde\sigma_1,A^1)\rangle\\
&\equiv & \mbox{tr}((L^{-1}\tilde\sigma_2^a L)^{\dag}
\;L^{-1}\tilde\sigma_1^b L)\;\hat{q}_{ab}+  
\mbox{tr}((L^{-1}A^2_a L)^{\dag}\; L^{-1} A^{1}_b L)\;\hat{q}^{ab},
\end{eqnarray*}
where $\hat{q}^{ab}$ is the metric constructed out of
$\hat{\sigma}^a$, and therefore real. We shall denote with a
$(\;\hat{}\;)$
tensors which are constructed with $\hat{\sigma}$. Since they are
${\bf SL}(2,{\bf C})$ scalars they are also the ones constructed with
$L\hat{\sigma}^a L^{-1}$. On the extended equations we shall raise and
lower indices with $\hat{q}^{ab}$ and its inverse.

Again, we consider just the case $N^a=0$, for the part
proportional to it is diagonal, hence symmetric, without any
conditions on $\tilde{\sigma}^a$.
Recalling the expressions for the principal symbols,

\begin{eqnarray}
\label{eq:op(sigma)2}
P_\sigma(\tilde\sigma,ik_a)\tilde\sigma_1{}^b&=& -
\frac{k_a}{\sigma }\;\;\hbox
{${}_{{}_{{}_{\widetilde{}}}}$\kern-.5em\rm N}
\mbox{tr}(\tilde{\sigma}_1{}^b\tilde{\sigma}^e)\eps^{a}{}_{ec}\tilde{\sigma}^c,
\end{eqnarray}

\begin{eqnarray}
\label{eq:op(A)2}
P_A(\tilde\sigma,ik_c)\;A^1{}_b&=& \frac{2}{\sigma}\;\hbox {${}_{{}_{{}_{\widetilde{}}}}$\kern-.5em\rm N} k_{[a} 
\mbox{tr}(A^1_{b]}\tilde{\sigma}_e)\eps^{ae}{}_c\tilde{\sigma}^c\nonumber\\
&&+\;\hbox {${}_{{}_{{}_{\widetilde{}}}}$\kern-.5em\rm N} \eps^{cde} \frac{k_c}{\sigma} \tilde\sigma_b\;\mbox{tr}(\tilde\sigma_e\;A^1_d)+\frac{2}{\sigma}\;\hbox {${}_{{}_{{}_{\widetilde{}}}}$\kern-.5em\rm N}\; \eps_b{}^{dc} \tilde\sigma_c\;\mbox{tr}(\tilde\sigma^a\; k_{[d}A^1_{a]}).
\end{eqnarray}

Once modified, they become
\begin{eqnarray}
\label{eq:op(hatsigma)}
\hat{P}_\sigma(\tilde\sigma,ik_a)\tilde\sigma_1{}^b
&=& - \frac{k_a}{\hat{\sigma}^4}\;\;\hbox
{${}_{{}_{{}_{\widetilde{}}}}$\kern-.5em\rm N}
\mbox{tr}(\tilde{\sigma}^b_{1}L\hat{\sigma}^eL^{-1})\hat{\eps}^{a}{}_{ec}L
\hat{\sigma}^cL^{-1},
\end{eqnarray}
and
\begin{eqnarray}
\label{eq:op(hatA)}
\hat P_A(\tilde\sigma,ik_c)\;A^1{}_b&=& \frac{2}{\hat\sigma}\;\hbox
{${}_{{}_{{}_{\widetilde{}}}}$\kern-.5em\rm N} k_{[a} 
\mbox{tr}(A^1_{b]}L\hat{\sigma}_eL^{-1})\hat\eps^{ae}{}_cL\hat{\sigma}^cL^{-1}\nonumber\\
&&+\;\hbox {${}_{{}_{{}_{\widetilde{}}}}$\kern-.5em\rm N}\hat \eps^{cde}
\frac{k_c}{\hat\sigma} L\hat\sigma_bL^{-1}\;\mbox{tr}(L\hat\sigma_e
L^{-1}\;A^1_d)+\frac{2}{\hat\sigma}\;\hbox
{${}_{{}_{{}_{\widetilde{}}}}$\kern-.5em\rm N}\; \hat\eps_b{}^{dc}
L\hat\sigma_c L^{-1}\;\nonumber \\
&&\times\mbox{tr}(L\hat\sigma^a L^{-1}\; k_{[d}A^1_{a]}).
\end{eqnarray}

Which is identical to the one used in the previous section, but now 
$\hat{\sigma}^c$ is by construction anti-hermitian even outside the 
reality condition sub-manifold, therefore $L\hat\sigma^a L^{-1}$ can 
be trivially rotated into an anti-hermitian one, 
and so symmetric-hyperbolicity follows as before. 
With this modification the standard proof of well posedness
of symmetric-hyperbolic systems apply.

\noindent {\bf Main Result}: {\it If
we assume   that data, $(\sigma_0^a,A^0_a)$, is given at a initial
surface $\Sigma_0$ such that it  belongs (locally) to the Sobolev
space $H^s(\Sigma_0)\times H^{s-1}(\Sigma_0)$, $s \geq 3$, and such
that $\tilde{q}^{ab}$ and its first time derivative at $\Sigma_0$ are
very close to be real, (that is,  we start with initial data in a 
sufficiently small neighborhood of the reality sub-manifold) then a 
local solution exists and it stays in these spaces along the generated
foliation.}

%%%%%%%%%%%%%%%%%%%%%%%%%%%%%%%%%%%%%%%%%%%%%%%%%%%%%%%%%%%%%%%%%%%%%%%%%%%

\section{Characteristic Directions and Strong Hyperbolicity}
\label{sec:CD}
%%%%%%%%%%%%%%%%%%%%%%%%%%%%%%%%%%%%%%%%%%%%%%%%%%%%%%%%%%%%%%%%%%%%%%%%%%%

It is of interest to see whether there are simpler modifications to the
equation system outside
the constraint sub-manifold which would give us well posedness
(strong-hyperbolicity) but with a
system not necessarily symmetric-hyperbolic. 
In doing this we shall calculate the complete set of
eigenvectors with their respective eigenvalues.  

Recalling the expressions for the principal symbols of the system,
(\ref{eq:op(sigma)2}), (\ref{eq:op(A)2}), we see that it is convenient to
use at   each point an
orthogonal triad $\{k_a, m_a^+,  m_a^-\}$ 
since we want to extend the results to a neighborhood of the reality
condition 
sub-manifold. Let  $k^a$ --the wave vector--  be real, with 
$|k_ak^a|=1$ and let   the other two vectors are taken to be null and orthogonal to $k_a$.
They are normalized such that, 
$m_a^+ m_b^- q^{ab} := k^{-1}:= (\sqrt{k_ak^a})^{-1}$,
and chosen so that when the metric becomes real
$\overline{m}_a^+ = m_a^-$ (complex null vectors). 
Note in particular that we have the relation:
$$
\eps_{abc}=i6k_{[a} m^+_b m^-_{c]}
$$

Thus the symbol can be diagonalized
in blocks with the following eigenvalues
$$
\lambda_0=i\;k_a N^a \qquad 
\lambda_+=i(\;k_a N^a+\;\hbox
{${}_{{}_{{}_{\widetilde{}}}}$\kern-.5em\it N}k\;\hbox
{${}_{{}_o}$\kern-.6em $\sigma $})\qquad 
\lambda_-=i(\;k_a
N^a-\;\hbox {${}_{{}_{{}_{\widetilde{}}}}$\kern-.5em\it N}k\;\hbox
{${}_{{}_o}$\kern-.6em $\sigma $})
$$

and the subspaces associated with the above eigenvalues are
\begin{eqnarray*}
E_{\sigma }{}^0&=&\mbox{Span}\{k_a\;k_d\;\hbox
{${}_{{}_o}$\kern-.6em
$\tilde\sigma $}^d, m^+_a\;k_d\;\hbox {${}_{{}_o}$\kern-.6em
$\tilde\sigma $}^d, m^-_a\;k_d\;\hbox {${}_{{}_o}$\kern-.6em
$\tilde\sigma $}^d\},\\
E_{\sigma }{}^+&=&\mbox{Span}\{k_a\; m^-_d\;\hbox
{${}_{{}_o}$\kern-.6em
$\tilde\sigma $}^d, m^+_a\;m^-_d\;\hbox {${}_{{}_o}$\kern-.6em
$\tilde\sigma $}^d,  m^-_a\; m^-_d\;\hbox {${}_{{}_o}$\kern-.6em
$\tilde\sigma $}^d\},\\
E_{\sigma }{}^-&=&\mbox{Span}\{k_a\;m^+_d\;\hbox
{${}_{{}_o}$\kern-.6em
$\tilde\sigma $}^d, m^+_a\;m^+_d\;\hbox {${}_{{}_o}$\kern-.6em
$\tilde\sigma $}^d, m^-_a\;m^+_d\;\hbox {${}_{{}_o}$\kern-.6em
$\tilde\sigma $}^d\},\\
E_{A }{}^0&=&\mbox{Span}\{k_a\;k_d\;\hbox {${}_{{}_o}$\kern-.6em
$\tilde\sigma $}^d, k_a\;m^+_d\;\hbox {${}_{{}_o}$\kern-.6em
$\tilde\sigma $}^d, k_a\;m^-_d\;\hbox {${}_{{}_o}$\kern-.6em
$\tilde\sigma $}^d\},\\
E_{A}{}^+&=&\mbox{Span}\{m^+_a\;k_d\;\hbox
{${}_{{}_o}$\kern-.6em
$\tilde\sigma $}^d, m^+_a\; m^+_d\;\hbox {${}_{{}_o}$\kern-.6em
$\tilde\sigma $}^d, m^+_a\;m^-_d\;\hbox {${}_{{}_o}$\kern-.6em
$\tilde\sigma $}^d\},\\
E_{A}{}^-&=&\mbox{Span}\{ m^-_a\;k_d\;\hbox
{${}_{{}_o}$\kern-.6em
$\tilde\sigma $}^d, m^-_a\;m^+_d\;\hbox {${}_{{}_o}$\kern-.6em
$\tilde\sigma $}^d, m^-_a\;m^-_d\;\hbox {${}_{{}_o}$\kern-.6em
$\tilde\sigma $}^d\},
\end{eqnarray*}

In order to clarify how we have computed these eigenvectors and eigenvalues, 
let us check  this result for   $A^1_b=m^+_bl_dk^d\;\in E_A^+$, with 
$l_a \in \{k_a, m_a^+,  m_a^-\}$. Consider the case     $N^a=0$, for the part proportional to it is diagonal, then
\begin{eqnarray}
&& P_A(\tilde\sigma,ik_c)\;A^1{}_b= \frac{2}{\sigma}\;\hbox
{${}_{{}_{{}_{\widetilde{}}}}$\kern-.5em\rm N} k_{[a} 
\mbox{tr}(A^1_{b]}\tilde{\sigma}_e)\eps^{ae}{}_c\tilde{\sigma}^c\nonumber\\
&&+\;\hbox {${}_{{}_{{}_{\widetilde{}}}}$\kern-.5em\rm N} \eps^{cde}
\frac{k_c}{\sigma}
\tilde\sigma_b\;\mbox{tr}(\tilde\sigma_e\;A^1_d)+\frac{2}{\sigma}\;\hbox
{${}_{{}_{{}_{\widetilde{}}}}$\kern-.5em\rm N}\; \eps_b{}^{dc}
\tilde\sigma_c\;\mbox{tr}(\tilde\sigma^a\; k_{[d}A^1_{a]})\nonumber\\
&&=\sigma\;\hbox
{${}_{{}_{{}_{\widetilde{}}}}$\kern-.5em\rm N}\;\left(2k_{[b}m_{a]}^+l_e\eps^{ae}{}_c\tilde \sigma^c+ 2k_{[a}m_{d]}^+l^d\eps^{a}{}_{bc}\tilde \sigma^c -k_a m_d^+ l_e\eps^{ade}\tilde \sigma_b\right)\nonumber\\
&&=-\frac{ik}{2}\sigma\;\hbox
{${}_{{}_{{}_{\widetilde{}}}}$\kern-.5em\rm N}\;(2\epsilon_{ban}m^{+n}\epsilon^{ae}{}_cl_e\tilde\sigma^c+ 2\epsilon_{adn}m^{+n}\epsilon^{a}{}_{bc}l^d\tilde\sigma^c \nonumber \cr
&&-\epsilon_{adn}m^{+n}\epsilon^{ade}l_e\tilde\sigma_b)\nonumber\\
&&=ik\sigma\;\hbox
{${}_{{}_{{}_{\widetilde{}}}}$\kern-.5em\rm N}\;m^+_bl_dk^d.
\end{eqnarray}

Note that, assuming  the metric to be real and  choosing  the  space-like foliation  such that   $k_a$ becomes   the normal to a time-like hypersurface
and  the
shift vector tangent to it, then we see that for each
$\tilde{\sigma}^a$ and $A_a$, 
there are three incoming characteristics, the same number of outgoing (as
should be the
case for gauges respecting time-direction symmetry) and three
characteristics which 
move along the boundary.

Note that the eigenvectors span the whole space on the solution
manifold, furthermore they are smooth functions on all arguments for
$k_a \neq 0$.
Furthermore, if the metric is real, that is, if we are in the reality
condition 
sub-manifold, then the eigenvalues are purely imaginary, thus we have a
strongly-hyperbolic
system. 
Can one make a simpler flow regularization and get strongly
hyperbolicity (a weaker
condition than symmetric-hyperbolicity)? That is, can we get away with
the $L$ transformation?  From the form of the eigenvalues it is clear
that this is the case. Indeed, 
one can modify the system outside the reality condition sub-manifold in
such a way that the 
eigenvalues are purely imaginary also there, but since in general $k$ is
complex, and depends on
$k_a$, the needed modification transforms the differential equation
system into a pseudo-differential one. We do not pursue this further
because such a modification would not be practical for most
applications one envisions.

%%%%%%%%%%%%%%%%%%%%%%%%%%%%%%%%%%%%%%%%%%%%%%%%%%%%%%%%%%%%%%%%%%%%%%%%%%%%%%%%%%%%%%%%%%%%%%%%%

%%%%%%%%%%%%%%%%%%%%%%%%%%%%%%%%%%%%%%%%%%%%%%%%%%%%%%%%%%%%%%%%%%%%%%%%%%%%%%%%%%%%%%%%%%%%%%%%%

\section{The evolution of constraints and reality conditions}
\label{ev-const-real}

Once a suitable extended  evolution system for general relativity is
shown to 
be symmetric hyperbolic in a whole neighborhood of the constraint and
reality
conditions sub-manifold, then standard results can be applied and
different 
results on well posedness follow, in particular local evolution is
granted
if the initial data is smooth enough. For instance, for the specific system at hand, if the initial data
$(\tilde{\sigma}^a,A_a)$ is in $H^s(\Sigma_0) \times
H^{s-1}(\Sigma_0)$, $s \geq 3$, then the solution remains for a finite
time in the corresponding spaces along the generated foliation
$\Sigma_t$.~\footnote{For the local problem one takes (through the
choice of lapse-shift pairs) the foliation to be such that all the
constant
time surfaces coincide at
their boundaries, that is,   a {\sl lens shaped} domain of evolution.}

Given a solution in a foliation, we can identify via the lapse-shift
pair any 
point on $\Sigma_t$ for any $t$ with a point at the initial surface,
namely the  points 
that lying in the integral curve of the four dimensional vector field,
$t^a = Nn^a+N^a$ and so we can pull back to the initial surface the pair 
$(\tilde{\sigma}^a,A_a)$, thus in the initial surface a solution can be
seen as a one 
parameter family of fields in $H^s(\Sigma_0) \times H^{s-1}(\Sigma_0)$. 

The following geometrical picture of evolution emerges: we have an infinite 
dimensional manifold, $K$, of pairs of soldering forms and connections
in a 
three dimensional manifold, $\Sigma_0$ belonging to $H^s(\Sigma_0)
\times H^{s-1}(\Sigma_0)$.
Given any lapse-shift pair we can generate an evolution, that after the
pull back 
is just a one parameter family of pairs in $K$, that is, a curve on that
manifold. 
The tangent vector to that integral curve is our twice  modified
evolution 
equation system. Of course not all these integral curves are solutions
to 
Einstein equations, for they would not generally satisfy the constraint
nor
the reality conditions, which in fact form a sub-manifold of $K$, denoted $P$.

 The relevant question is   whether the integral curves
which
start at $P$ remain on $P$, for they conform the true solutions to
Einstein's 
equations. This would happens if and only if the tangent vector fields
to the
integral curves are themselves  tangent to the sub-manifold $P$. 
In order to see that   one can proceed as follows:

We let $RE$, and $CE$  denote the reality conditions {\sl
quantities},
namely $\Im q^{ab}$, $\Im \pi_{ab}$, and the constraint equations {\sl
quantities},
namely $\tilde{C}$, $C$, and $C^a$ respectively.~\footnote{This is basically a
coordinatization 
of a neighborhood of $P$ in $K$.} 
We smear these expressions out with smooth tensors and so obtain
maps from the manifold $K$ into the complex numbers,
\begin{eqnarray}
  \label{eq:esm.constraints}
  RE_{f_{ab}}{(\tilde{\sigma},A)} &:=& \int_\Sigma f_{ab} \Im q^{ab} \;\mbox{d}x^3\\
  RE_{f^{ab}}{(\tilde{\sigma},A)} &:=& \int_\Sigma f^{ab} \Im \pi_{ab} \;\mbox{d}x^3\\
  CE_{\tilde{f}}{(\tilde{\sigma},A)} &:=& -i\sqrt{2}\int_\Sigma \mbox{tr} (\tilde{f}
\tilde{C}) \;\mbox{d}x^3\\
  CE_{f}{(\tilde{\sigma},A)} &:=&-i\sqrt{2} \int_\Sigma f C\;\mbox{d}x^3 \\
  CE_{f_{a}}{(\tilde{\sigma},A)} &:=& -i\sqrt{2}\int_\Sigma f_{a} C^a\;\mbox{d}x^3, 
\end{eqnarray}
We thus see that the reality-conditions-constraint sub-manifold $P$ can
be
defined as the intersection of zero level set of each of this infinite
set of maps. If these maps are sufficiently differentiable so that
their
gradients are well defined,   it is clear that the tangency 
condition of the evolution vector field is just the requirement that
when   contracted with the differentials of the above defined maps
the result should vanish at points of $P$. But these contractions are
just the smeared out version of the time derivatives of the reality
and constraint equations. Thus, provided we have enough
differentiability,\footnote{Note that in this manipulations no surface
term
arises, for in treating the local problem the lapse-shift pair vanishes
at the 
boundary} we only have to check that this time derivatives
vanish at $P$. 

Since these calculations need only be done at points of $P$, it only
involves the original evolution vectors field, and so the standard
results can be used, namely that the time derivative of the constraint
equations, and the time derivative of the reality conditions vanish at
$P$. The first result  (see  appendix\ref{evolconstr})  follows from
the constraint algebra calculated in \cite{ash}. The second one,
basically, follows from the fact
 that the Ashtekar system is equivalent, up to terms proportional to
the constraints, to Einstein's evolution equations. Since in that
equivalence one does not use any hermiticity nor reality condition
explicitly, Ashtekar's equations are in fact equivalent to {\it
complexified} gravity that is to equations identical to Einstein's, but
where the metric and the second fundamental form on each slice can be
complex. Thus, at points where both the metric and the second
fundamental form are real, and the constraints are satisfied, that is,
at $P$, and provided that the three metric is invertible, the imaginary
part of these tensors clearly vanishes. We have also verified this
directly (see appendix \ref{real}, and see also previous works, \cite{ash-rom} and \cite{immirzi}), obtaining
\begin{eqnarray}
\label{eq:realcond}
{\cal L}_t  q_{ab}&=&{\cal L}_{\vec N}q_{ab}-2N\pi_{(ab)}\nonumber\\
{\cal L}_t \pi_{ab}&=&{\cal L}_{\vec
N}\pi_{ab}+N\pi\pi_{ab}-2N\pi_a{}^e\pi_{eb}-N
R_{ab}-D_aD_bN\nonumber\\
&&-\frac{N}{2}q_{ab}\left(R+\pi^2-\pi_{dc}\pi^{cd}\right)-iN\epsilon_{ab}{}^d D^c\left(\pi_{dc}-\pi
q_{dc}\right)\nonumber\\
&&+2N\left(\pi^d{}_a\pi_{[db]}-\pi\pi_{[ab]}+\pi_{[ad]}\pi^d{}_b\right)-iN\epsilon_b{}^{dm} D_a\pi_{[dm]}\nonumber \\
&&-\frac{i}{2}q_{ab}\epsilon^{cmd}\left(D_cN\pi_{[md]}-ND_c\pi_{[md]}\right)
\end{eqnarray}
where
$$
\pi_{ab}=-\mbox{tr}(\pi_a\sigma_b)\quad\mbox{and}\quad\pi_{[ab]}=
-\frac{i\epsilon_{abc}}{2}\;\mbox{tr}(\sigma^c{\cal D}_d\sigma^d).
$$
Since the terms, which could give an imaginary contribution to the
expression when $q_{ab}$ and
$\pi_{ab}$ are real, are proportional to the constraints, we see that at
points of $P$ the 
evolution of the imaginary parts $q_{ab}$, and $\pi_{ab}$ vanishes. Note
that the real part of $\pi_{ab}$ corresponds to the extrinsic curvature,
i.e. $\pi_{(ab)}=K_{ab}$. We conclude that, 

\noindent {\bf Main Result}:  {\it Initial data satisfying both the reality
conditions, and the constraint equations  have
an evolution which stays inside $P$, and so are solutions to the real
and complete set of Einstein's equations}.~\footnote{The above argument is only valid for establishing local well
posedness. For   initial-boundary-value problems, one must be aware of
boundary terms and impose conditions for them to also vanish.}

We will conclude with a brief discussion of some issues relevant to numerical relativity.

What does this mean when one does not solve Einstein's equations
exactly, but just approximate them via numerical simulations or other
means? If the approximations were a contraction map as the one often
used to prove existence of exact solutions of symmetric hyperbolic
systems,
then it would follow that the approximate solution must approach the
manifold $P$ as it is refined and made closer to the exact solution.
In practice the approximation schemes are not contractive as required.
 Even after refinement, if   the method yields  convergence to the manifold $P$, 
this convergence could be very slow.

To explore this problem, and in analogy with a similar work of Fritelli\newline ~\cite{simo}, we have looked at the evolution equations that the 
constraints satisfy when the flow is extended , but assuming that the
reality
conditions hold.  
From this study we have found that the constraint quantities satisfy
by
themselves a symmetric hyperbolic system of equations. 
Thus, initial data sets which satisfy exactly the reality conditions,
but are just near the constraint sub-manifold, if evolved by a scheme
that respects the reality conditions then  would stay ``near'' the
constraint
sub-manifold, in the sense that their departure would be bounded by a
constant
that depends only on the Sobolev\newpage
\noindent norm of the initial data set. 
Unfortunately symmetric hyperbolicity by itself does not allow 
a finer control on the bound, which in principle could, even for linear
equations,
grow exponentially with the evolution time. Thus, this crude bound
is not enough for controlling numerical simulations, but its absence  would certainly make simulations very hard to implement.
 
 The evolution equations for the constraints have been calculated in
appendix \ref{evolconstr}. In order to prove  the symmetric
hyperbolicity we  just need to consider the  principal part of this set
\begin{eqnarray}
\label{eq:constraintevol2}
\dot{\tilde C}&=&N^a\partial_a \tilde C-\frac{i}{\sqrt{2}} \;\hbox
{${}_{{}_{{}_{\widetilde{}}}}$\kern-.5em\it
N}\;[\tilde\sigma^a,\partial_a \tilde C],\nonumber\\
\dot{ C}&=&N^a\partial_a  C- i\sqrt{2} \;\hbox
{${}_{{}_{{}_{\widetilde{}}}}$\kern-.5em\it N} \sigma^2 \partial^a
C_a,\\
\dot{ C_a}&=&N^b\partial_b  C_a+\frac{i}{\sigma^2}\;\hbox
{${}_{{}_{{}_{\widetilde{}}}}$\kern-.5em\it N}\;\tilde
\epsilon_a{}^{bc}\partial_b C_c+\frac{i}{\sqrt{2}}\;\hbox
{${}_{{}_{{}_{\widetilde{}}}}$\kern-.5em\it N}\;\partial_a C.\nonumber
\end{eqnarray}
 Then we prove the following,

\noindent{\bf Lemma III.1:} {\em The equation system
(\ref{eq:constraintevol2})
  for any fixed, but arbitrary lapse and shift fields is a symmetric
  hyperbolic system in the reality condition sub-manifold}.

\noindent{{\bf Proof:}}

The eigenvectors and the eigenvalues can now be easily calculated,
note that the principal symbol in this system
\begin{equation}
\label{eq:ps1}
P(\;\hbox {${}_{{}_o}$\kern-.6em\rm u},ik_a)=P_{\tilde C}(\;\hbox
{${}_{{}_o}$\kern-.6em\rm u},ik_a) \oplus P_{ \vec C}(\;\hbox
{${}_{{}_o}$\kern-.6em\rm u},ik_a),
\end{equation}
here $\hbox {${}_{{}_o}$\kern-.6em\rm u}=(\hbox {${}_{{}_o}$\kern-.6em
  $\tilde \sigma$},\hbox {${}_{{}_o}$\kern-.6em \it A})$, and $\vec
C=(C,C_b)$.

As above we use the orthogonal triad $\{k^a, m^a, \overline{m}^a\}$ being
$m^a$ and $\overline{m}^a$ null vectors. Thus the symbol can be diagonalized
in blocks with the following eigenvalues
$$
\lambda_0=i\;k_a N^a \qquad \lambda_{-}=i(\;k_a N^a-\;\hbox
{${}_{{}_{{}_{\widetilde{}}}}$\kern-.5em\it N}\;\hbox
{${}_{{}_o}$\kern-.6em $\tilde\sigma $})\qquad \lambda_{+}=i(\;k_a
N^a+\;\hbox {${}_{{}_{{}_{\widetilde{}}}}$\kern-.5em\it N}\;\hbox
{${}_{{}_o}$\kern-.6em $\tilde\sigma $}),
$$
and the subspaces associated with the above eigenvalues are
\begin{eqnarray*}
E_{\tilde C }{}^0&=&\mbox{Span}\{k_d\;\hbox
{${}_{{}_o}$\kern-.6em
$\tilde\sigma $}^d\},\qquad
E_{\tilde C }{}^-=\mbox{Span}\{\;m_d\;\hbox
{${}_{{}_o}$\kern-.6em
$\tilde\sigma $}^d\},\qquad
E_{\tilde C }{}^+=\mbox{Span}\{\overline{m}_d\;\hbox
{${}_{{}_o}$\kern-.6em
$\tilde\sigma $}^d \},\\
E_{\vec C }{}^+&=&\mbox{Span}\{ (0, m_a), (\;\hbox
{${}_{{}_o}$\kern-.6em
$\tilde\sigma $}, \frac{i}{\sqrt 2}k_a )\},\\
E_{\vec C }{}^-&=&\mbox{Span}\{(0,\overline{m}_a), (\;\hbox
{${}_{{}_o}$\kern-.6em
$\tilde\sigma $}, -\frac{i}{\sqrt 2}k_a )\}.
\end{eqnarray*}

This set is a complete orthonormal set of eigenvectors with respect to
the inner product
\begin{eqnarray*}
\langle u^2, u^1\rangle &\equiv &\langle(\tilde C^2,\vec C^2),(\tilde
C^1,\vec C^1)\rangle\\
&\equiv & \mbox{tr}(\tilde C^{\dag 2} \tilde C^1) +  \frac{1}{2 \;\hbox
{${}_{{}_o}$\kern-.6em $\tilde\sigma $}^2}C^{2\dag}
C^{1}+q^{ab}C_a^{2\dag} C_b^1,
\end{eqnarray*}
and since the eigenvalues are purely imaginary the principal symbol is
anti-hermitian. This concludes the proof of the Lemma.$\spadesuit$

%%%%%%%%%%%%%%%%%%%%%%%%%%%%%%%%%%%%%%%%%%%%%%%%%%%%%%%%%%%%%%%%%%%%%%%%%%

\section{Conclusions}

We have studied several aspects of Einstein's evolution equations as
given
in Ashtekar's formalism. We first have shown that when the reality
conditions are satisfied, that is, when there is a ${\bf SL}(2,{\bf C})$
transformation
that rotates the soldering form into a anti-hermitian one, the system is
symmetric
hyperbolic. 

Contrary to what one might expect at first , this condition is not enough to
conclude that there are solutions, even local ones, to the evolution
equations.
The problem being that, since the reality conditions are not linear
conditions on the variables on which the problem is formulated, the
usual
contractive map of successive approximations does not respect them, and
so evolution  
occurs in this approximations along paths where the system
is not symmetric-hyperbolic, meaning that they can not be appropriately bounded by the usual methods.

To remedy this problem we have proposed two further modifications.  One of the modifications, called the {\em regularized flows}  is a standard trick commonly
used to grant symmetric hyperbolicity, suitably extended  to the case
under consideration, that is, to account for the intermediate ${\bf
SL}(2,{\bf C})$
rotation mentioned above. In that case one obtains a symmetric
hyperbolic system of equations {\sl even outside the reality condition
sub-manifold}. This allows then to establish the local well posedness of the problem by
standard 
procedures.

The second modification has the advantage of keeping
unmodified the eigenvalues-eigenvectors structure of the principal part, 
which have a simple, and therefore
probably useful, expression in terms of the solution, 
at the expense of transforming the
differential equations into non local ones, namely into
pseudo-differential
equations. This modification can be implemented in numerical schemes
--using fast Fourier transform-- but only for a limited type of
boundary conditions.

After the regularization of the flow and the subsequent implication of
the existence of solutions of given differentiability, we discuss the
problem
of making sure that  the solution, whose local existence we are asserting, does
satisfy the whole set of Einstein's equations. In other words, whether
integral
curves to the evolution vector fields starting at the 
reality-constraint sub-manifold stay there.
Since we have already shown that the solutions are unique (if sufficiently
differentiability is assumed) then it is enough to show tangency of
the vector field with respect to that sub-manifold. Provided
some
differentiability conditions are satisfied, this is equivalent to
standard
results which were nevertheless reviewed.

We concluded section \ref{ev-const-real}
with   some considerations about the problem
of numerical simulations, or approximations in general. Since errors
are unavoidable in numerical algorithms one can at most consider
evolution in a hopefully
small neighborhood of the reality-condition-constraint sub-manifold.
Thus,
even when --in the best of the cases-- knowing that eventually, by
refinements of the approximation scheme,  the solution would converge
to that sub-manifold, one would like to have a priori estimates of the
deviation as a function of the initial error.

 We have shown, following
the ideas of Fritelli for the ADM formulation, that the constraint
system   --while at the reality sub-manifold-- satisfies by itself a set
of
symmetric hyperbolic equations. Thus such an a priori bound follows
directly. We consider this a preliminary result, for the bounds that
follow from symmetric hyperbolicity are too crude for numerical
purposes,
and so a further refinement, this time using information about the
full system of equations (contrary to symmetric-hyperbolicity which only
uses
the principal part structure of them), seems to be needed.

We will conclude with a remark. A substantial improvement on the handling of these equations would be
the possibility of imposing reality conditions in a linear way, that is,
by requiring the hermiticity (or anti-hermiticity if considered as
matrices) of the soldering forms. Unfortunately with the present
scheme
of modifying the equations outside the constraint sub-manifold, the
hermiticity condition restricts unnecessarily the possible evolutions,
for it basically fixes a unique lapse. Otherwise the system can not be made
symmetric hyperbolic in an straightforward way. 
This restriction is unnatural, for one would expect that  this condition
could be enforced fixing the ${\bf SU}(2)$ gauge freedom, and not the
one associated with diffeomorphisms. But all attempts to fix the ${\bf SU}(2)$ gauge appropriately seems to involve (at best) elliptic
conditions on the "time 
 component" of the connection.  Is there an alternative avenue to handle the linear reality conditions?

%%%%%%%%%%%%%%%%%%%%%%%%%%%%%%%%%%%%%%%%%%%%%%%%%%%%%%%%%%%%%%%%%%%%%%%%%%
\subsection*{Acknowledgments}
We are  indebted
to A. Ashtekar,  H. O. Kreiss and H. Friedrich for many enlightening conversations
and suggestions. 

%%%%%%%%%%%%%%%%%%%%%%%%%%%%%%%%%%%%%%%%%%%%%%%%%%%%%%%%%%%%%%%%%%%%%%%%%%

%\appendix 
\begin{center}
{\bf APPENDIX A: THE CONSTRUCTION OF THE ${\bf SL}(2,{\bf C})$ TRANSFORMATION.}
\label{sec:L}
\end{center}
Given a soldering form, $\sigma^a$, we shall define here a procedure to 
construct a ${\bf SL}(2,{\bf C})$ transformation such that, when the
metric is real the
transformed soldering form is anti-hermitian. 
Given a transformation $L$ making $\tilde{\sigma}^a$ anti-hermitian,
there is a whole set of them with the same property, namely one can
left-right multiply the new one by a ${\bf SU}(2)$ rotation and obtain a
new
one with the same anti-hermiticity property. To remove this
arbitrariness
in our construction we fix a real basis (not orthonormal), $\{e^a_i\}$
$i=1,2,3$ of vector fields. For simplicity we shall work here with
frames,$\{E^a_i\}$ $i=1,2,3$ so that $\sigma^a = E^a_i\tau^i$ where
$\tau^i$ is an anti-hermitian Pauli set of matrices. We then have
$E^a_iE^b_j\delta^{ij} = q^{ab}$, and so
$E^a_iE^b_jq_{ab}=\delta_{ij}$. 
We define $L$ as the transformation such that the transformed
frame $\{\hat{E}^a_i\}$ $i=1,2,3$ satisfies:
\begin{eqnarray}
\hat{E}^a_1 &=& b^1_1 e^a_1  \nonumber \\
\hat{E}^a_2 &=& b^1_2 e^a_1 + b^2_2 e^a_2 + b^3_2 e^a_3 \;\;\;
\mbox{with}\;\; \frac{b^1_2}{b^2_2}, \;
\frac{b^3_2}{b^2_2} \;\; \mbox{real.} \nonumber
\end{eqnarray}
The ortho-normality condition then fixes the third vector, and so the
complete
transformation. Notice that
if the initial frame can be rotated to a real one, that is, if the
metric is real, then the rotated frame we are constructing will be
real.

We let $\hat{E}^a_i = C_i{}^je^a_j$ and $\hat{E}^a_i = L_i{}^jE^a_j =
L_i{}^jC_j{}^ke^a_k = {\cal L}_i{}^je^a_k$.  Since $C_i{}^j$ is
invertible, knowing ${\cal L}$ is equivalent to knowing $L$, we shall
construct ${\cal L}$.

We have, ${\cal L}_1{}^1 = b^1_1$, ${\cal L}_1{}^2 = {\cal L}_1{}^3 =
0$, $b^1_1$ is fixed by the ortho-normality condition: $1 = (b^1_1)^2
e_1\cdot e_1$, we choose the root with positive real part.

Next we have ${\cal L}_2{}^1 = b^1_2$ ${\cal L}_2{}^2 = b^2_2$ ${\cal
  L}_2{}^3 = b^3_2$, and two conditions, the normality condition,
$$
1 = (b^1_2)^2 e_1 \cdot e_1 + (b_2^2)^2 e_2 \cdot e_2 + (b_2^3)^2
e_3 \cdot e_3 + 2 b^1_2 b^2_2 e_1 \cdot e_2 + 2 b_2^2 b^3_2 e_2 \cdot
e_3 + 2 b_2^1 b_2^3 e_1 \cdot e_3.
$$
and the orthogonality one,
$$
0 = b^1_1(b_2^1 e_1 \cdot e_1 + b^2_2 e_1 \cdot e_2 + b_2^3 e_1
\cdot e_3)
$$

We require that $\Im(\frac{e_1 \cdot e_3}{e_1 \cdot e_1}) \neq 0$,
otherwise
we impose the reality condition to $b^1_2/b^3_2$ and take $b^2_2 = 0$.
We multiply the orthogonality relation by a phase $e^{i\theta}$
such that $\Im(e^{i\theta} e_1\cdot e_2) = 0$, we then divide by
$b^2_2$ and assume the quotient of the $b$'s to be real. Taking real
and imaginary part we get two real linear equations:
$$
\frac{b^1_2}{b^2_2} \Im(e^{i\theta}e_1\cdot e_1) +
\frac{b^3_2}{b^2_2} \Im(e^{i\theta}e_1\cdot e_3) = 0
$$
$$
\frac{b^1_2}{b^2_2} \Re(e^{i\theta}e_1\cdot e_1) +
\frac{b^3_2}{b^2_2} \Re(e^{i\theta}e_1\cdot e_3) = - |e_1\cdot e_2|,
$$
which, because of the requirement above, have a non-vanishing
determinant.
Thus, it has a real solution for the quotients. We then solve the
normality condition for $b^2_2$, choosing, as before, the root with
the positive real part. The remaining coefficients of ${\cal L}$ are
completely determined by the remaining ortho-normality conditions
(three equations) and the condition that the real part of $b^3_3$ be
positive.

%%%%%%%%%%%%%%%%%%%%%%%%%%%%%%%%%%%%%%%%%%%%%%%%%%%%%%%%%%%%%%%%%%%%%%%%%%
%\appendix
\begin{center}
{\bf APPENDIX B: BRIEF OVERVIEW OF THE HAMILTONIAN FORMULATION IN ASHTEKAR'S VARIABLES} \end{center}
\label{sec:ash}
In this section we shall present a brief overview of the Hamiltonian
formulation in spinorial variables in order to get Einstein field
equations as a system of evolution and constraints equations and
thereby to study the evolution of the constraints (we follow
\cite{ash} and \cite{ash-rom}).
\setcounter{section}{0}
\section{ The Lagrangian Framework}
Let us consider a four-manifold $M$ which has topology $\Sigma \times
{\bf R}$, for some three-manifold $\Sigma$, with a four dimensional
${\bf SL}(2,{\bf C})$ soldering form, $\sigma^a{}_A{}^{A'}$, and a
connection ${}^4A_a{}_A{}^{B}$ which acts on the unprimed spinor
indices. We restrict the soldering form to be anti-Hermitian so that
it defines a real space-time metric via
$g^{ab}=\sigma^a_{AA'}\sigma^{bAA'}$ with signature $(-+++)$ and a
unique torsion-free derivative operator $\nabla$ compatible with
$\sigma^a{}_A{}^{A'}$. Finally, the derivative operator ${}^4{\cal
  D}_a$ defined by ${}^4A_a{}_A{}^{B}$ via ${}^4{\cal
  D}_a\lambda_A=\partial_a \lambda_A+ {}^4A_a{}_A{}^{B}\lambda_B$ acts
on unprimed spinors.

The gravitational part of the Lagrangian density of weight 1 is:

$$
{\cal L}=({}^4\sigma)\sigma^a{}_A{}^{A'} \sigma^b{}_{BA'}
{}^4F_{ab}{}^{AB},
$$
where $({}^4\sigma)$ is the determinant of the inverse soldering
form and ${}^4F_{ab}{}_{A}{}^B$ is the curvature tensor of
${}^4A_a{}_A{}^{B}$.

It is useful to  perform a $3+1$ decomposition of the action, 
this shall be needed latter in order to pass on to the Hamiltonian 
framework.

Let us introduce on $M$ a smooth function $t$ whose gradient is
nowhere vanishing and whose level surfaces $\Sigma_t$ are each
diffeomorphic to $\Sigma$. Let $t^a$ be a smooth vector field on $M$
with affine parameter $t$ and on each level surface, let $n^a$ be the
future-directed, unit,time-like vector field orthogonal to $\Sigma_t$.
Denote the induced positive-definite metric on $\Sigma_t$ by
$q_{ab}=g_{ab}+n_an_b$ and obtain the lapse and shift fields $N$ and
$N^a$ by projecting $t^a$ into and orthogonal to $\Sigma_t$ , i.e.
$t^a=Nn^a+N^a$.

Identifying unprimed ${\bf SL}(2,{\bf C})$ spinors on $M$ with ${\bf
  SU}(2)$ on $\Sigma_t$, we introduce the soldering form
$\sigma^a{}_A{}^B$ on ${\bf SU}(2)$ spinors. It defines the three
metric on $\Sigma_t$ as $q^{ab}=-\mbox{tr }(\sigma^a\sigma^b$ ). Here
a matrix notation is employed, and shall be used in the following, for
unprimed spinor indices in which adjacent summed indices go from upper
left to lower right, e.g., $(\sigma^a\sigma^b)_A{}^B=\sigma^a{}_A{}^C
\sigma^b_C{}^B$.

Let $A_a{}_A{}^B$, $F_{ab}{}_{A}{}^B$ and ${\cal D}_a $ (the {\it Sen
  connection}) be the pull-backs to $\Sigma_t$ of ${}^4A_a{}_A{}^{B}$,
${}^4F_{ab}{}_{A}{}^B$ and $ {}^4{\cal D}_b$ respectively.  Then ${\cal
  D}_a\lambda_A= \partial_a\lambda_A+ A_a{}_A{}^{B}\lambda_B$ and $
F_{ab}=2\partial_{[a}A_{b]}+[A_a,A_b].$

Finally, there is a natural (canonical) spinorial connection
associated with the three-metric such that $D_a \sigma^b=0$.  It
relates to the Sen connection via the extrinsic curvature $K_{ab}$on
$\Sigma_t$ as follows
\begin{eqnarray*}
{\cal D}_a\lambda_A&=&D_a\lambda_A+\frac{i}{\sqrt 2}
K_{aA}{}^B\lambda_B\\
&=&\partial_a\lambda_A+\Gamma_{aA}{}^B\lambda_B+\frac{i}{\sqrt 2}
K_{aA}{}^B\lambda_B,
\end{eqnarray*}
where $K_{aA}{}^B=K_{ab}\sigma^a{}_A{}^B$ and $\Gamma_{aA}{}^B$ is the
spin connection 1-form of $D$. Then $A_a=\Gamma_a+\frac{i}{\sqrt 2}
K_a$ and using the fact that the derivative $D$ is compatible with
$\sigma^a$, i.e.
$$
D_a\sigma_b=\partial_a\sigma_b-\Gamma_{ab}{}^c\sigma_c+[\Gamma_a,\sigma_b]=0
$$
with $\Gamma_{ab}{}^c$ denoting the Christoffel symbols; we
calculate
\begin{equation}
\label{eq:gamma}
\Gamma_a=\frac{\epsilon^{bcd}\sigma_a}{2\sqrt{2}}\mbox{tr
  }(\sigma_b\partial_c\sigma_d)-
\frac{\epsilon^{bcd}\sigma_b}{\sqrt{2}}\mbox{tr
  }(\sigma_a\partial_c\sigma_d).
\end{equation}
where the orientation three-form on $\Sigma_t$ is written as
$\epsilon^{abc}= -\sqrt{2}\mbox{tr }(\sigma^a\sigma^b\sigma^c)$.

The action \footnote{The surface terms are not included here.} can be
expressed in terms of only of three-dimensional fields:

$$
S=\int \mbox{dt}\int\mbox{d}^3x\;\mbox{tr}\bigg(\sqrt{2}\; i
\tilde\sigma^b({\cal L}_t A_b-{\cal D}_b({}^4A\cdot t)) -\;\hbox
{${}_{{}_{{}_{\widetilde{}}}}$\kern-.5em\it
  N}\;\;\tilde\sigma^a\tilde\sigma^b F_{ab}-\sqrt{2}\; i
N^a\tilde\sigma^b F_{ab}\bigg).
$$
where $\tilde\sigma^a{}_{AB}=(\sigma)\;\sigma^a{}_{AB}$, $\;\hbox
{${}_{{}_{{}_{\widetilde{}}}}$\kern-.5em\it N}=(\sigma)^{-1}\; N$ and
the Lie derivatives treat internal indices as scalars.

The action depends on five variables $\;\hbox
{${}_{{}_{{}_{\widetilde{}}}}$\kern-.5em\it N}$, $N^a$, ${}^4A\cdot
t$, $A_{aA}{}^B$ and $\tilde\sigma^a{}_{A}{}^{B}$. The first three
variables play the role of the Lagrange multipliers, only the last two
are dynamical variables.  Varying the action with respect to the
Lagrange multipliers we obtain the constraint equations:
\begin{eqnarray}
C(\tilde\sigma,A)&:=&\mbox{tr}(\tilde\sigma^a\tilde\sigma^bF_{ab})=0,\nonumber\\
C_a(\tilde\sigma,A)&:=&\mbox{tr}(\tilde\sigma^bF_{ab})=0,\label{eq:constraint}
\\
\tilde C_A{}^B(\tilde\sigma,A)&:=&{\cal D}_a
\tilde\sigma^a{}_A{}^B=0,\nonumber
\end{eqnarray}
and varying with respect to the dynamical variables, yields the
evolution equations:
\begin{eqnarray}
\label{eq:evolution0}
{\cal L}_t\tilde\sigma^b&=&-[{}^4A\cdot t,\tilde\sigma^b]+2{\cal
D}_a(N^{[a}\tilde\sigma^{b]})-\frac{i}{\sqrt{2}}{\cal D}_a(\;\hbox
{${}_{{}_{{}_{\widetilde{}}}}$\kern-.5em\it
N}[\tilde\sigma^{a},\tilde\sigma^{b}]),\nonumber\\
{\cal L}_t A_b&=&{\cal D}_b({}^4A\cdot
t)+N^aF_{ab}+\frac{i}{\sqrt{2}}\;\;\hbox
{${}_{{}_{{}_{\widetilde{}}}}$\kern-.5em\it N}[\tilde\sigma^a,F_{ab}].
\end{eqnarray}

\section{The Hamiltonian Framework}

Recall that the dynamics of a mechanical system can be achieved having
what is called a {\em symplectic manifold}, i.e a pair $(\Gamma,
\Omega_{\alpha\beta})$, where $\Gamma$ is an even-dimensional
manifold, and $\Omega_{\alpha\beta}$ a {\em symplectic form}, i.e a
2-form which is closed and nondegenerate. Given any function $f:\Gamma
\to {\bf R}$, the {\em Hamiltonian vector field} of $f$ is defined by
$ X_f{}^\alpha=\Omega^{\beta\alpha}\nabla_\beta f.  $ Given any vector
field $v^\alpha \in T_x\Gamma$, we say that $v^\alpha$ is a {\em
  symmetry} of the symplectic manifold if it leaves the symplectic
form invariant, i.e if $ {\cal L}_v\Omega_{\alpha\beta}=0, $ in which
case the diffeomorfisms generated by $v^\alpha$ are called {\em
  canonical transformations}.

Given two functions $f, g:\Gamma \to {\bf R}$, their {\em Poisson
  bracket} is defined by
$$
\{f,g\}:=\Omega^{\alpha\beta}\nabla_\alpha f \nabla_\beta g\equiv
{\cal L}_{X_f}g\equiv -{\cal L}_{X_g}f.
$$
Thus the dynamics of a physical system is given by assigning a {\em
  phase space} $(\Gamma, \Omega_{\alpha\beta}, H)$ on which the
evolution is generated by the Hamiltonian $H$ from the initial state.
Hence for any observable $f$
$$
\dot f = {\cal L}_{X_H} f=\{H,f\}.
$$
When the system is constrained, i.e. when there are points in the
phase space that can not be reached by the physical system, the system
remains in a sub-manifold called {\em constraint sub-manifold}, that can
be specified by the vanishing of a set of functions $ \bar\Gamma
=\{p\in \Gamma / C_i(p)=0, \mbox{for}\; i=1 \cdots m\}.  $ The
constrained system is said to be {\em first class} if there exist
functions $f_{ij}{}^k, i,j,k=1,\cdots , m$, called {\em structure
  functions}, such that $ \{C_i,C_j\}=f_{ij}{}^k C_k.  $

In the formulation of General relativity in Ashtekar variables, the
configuration space ${\cal C}$ is the space of all weighted soldering
forms $\tilde\sigma^a$, and the phase space $\Gamma$ is the cotangent
bundle over ${\cal C}$.  The phase space is represented by the pairs
$(\tilde\sigma^a, A_b)$ (these variables, apart from a numerical factor, the new canonically conjugate pair,
).  The action of a cotangent vector
$A_{a}$ on any tangent vector $(\delta\sigma)^a $ at a point
$\tilde\sigma^a $ of ${\cal C}$ is given by
$$
A(\delta\sigma)=-\int_\Sigma \mbox{d}^3x
\;\mbox{tr}(A_a\delta\sigma^a).
$$

We will not discuss boundary conditions here, but just remark that one
possible choice would be to require that the canonically conjugate
fields should admit a smooth extension to the point at spatial
infinity if the three-surface $\Sigma$ is made into the three sphere
by the one point compactification, for an extensive discussion of
fall-off properties of the fields see \cite{ash}.

Thus in order to obtain the constraint sub-manifold where the physical
gravitational states take place, we need to construct functionals from
the constraints (\ref{eq:constraint}), i.e. we need to smear out these
constraints with a function $\;\hbox
{${}_{{}_{{}_{\widetilde{}}}}$\kern-.5em\it N}$, a vector field $N^a$
and an anti-hermitian traceless $N_A{}^B$ test fields. We define
\begin{eqnarray}
C_N(\tilde\sigma,A)&:=&-i\;\sqrt{2} \int_\Sigma \mbox{d}^3x \quad
\mbox{tr}(N{\cal D}_a\tilde \sigma^a),\nonumber\\
C\;{}_{\hbox {${}_{{}_{{}_{\widetilde{}}}}$\kern-.5em {\small \it
N}}}(\tilde\sigma,A)&:=&-i\sqrt{2}\int_\Sigma \mbox{d}^3x \; \hbox
{${}_{{}_{{}_{\widetilde{}}}}$\kern-.5em\it N}\;\;
\mbox{tr}(\tilde\sigma^a\tilde\sigma^bF_{ab}),\label{eq:constraint1}
\\
C_{\vec N}(\tilde\sigma,A)&:=&-i\sqrt{2}\int_\Sigma \mbox{d}^3x \; N^a
\; \mbox{tr}(\tilde\sigma^bF_{ab}-A_a{\cal D}_b\tilde
\sigma^b),\nonumber
\end{eqnarray}
then we should ask if these constraint functions are {\em first class}
in the above terminology, to answer this question we have to calculate
their Poisson brackets.  For that purpose, we introduce the symplectic
form
$$
\Omega|_{(\sigma,A)}\bigg(( \delta\sigma,\delta A),(
\delta\sigma',\delta A')\bigg):=\frac{i}{\sqrt{2}}\int_\Sigma
\mbox{d}^3x \;\mbox{tr}(\delta A_a \delta\sigma'^a-\delta A'_a
\delta\sigma^a),
$$
where $( \delta\sigma,\delta A)$ and $( \delta\sigma',\delta A')$
are any two tangent vectors at the point $(\sigma,A)$. Thus given two
functionals $f$ and $g$ their Poisson bracket is
$$
\{f,g\}=\frac{i}{\sqrt{2}}\int_\Sigma \mbox{d}^3x\;
\mbox{tr}\left(\frac{\delta f}{\delta A_a}\frac{\delta g}{\delta
    \sigma^a}-\frac{\delta f}{\delta \sigma^a}\frac{\delta g}{\delta
    A_a}\right).
$$
The Poisson bracket between the constraint functionals and our
fundamental variables are:
\begin{eqnarray}
\{\tilde\sigma^a{}_{MN}, A_b{}^{AB}\}&=&-\frac{i}{\sqrt{2}}\delta (x,y)
\delta_M{}^{(A}\delta_N{}^{B)},\nonumber\\
\{C_N,\tilde\sigma^a \}&=&[\tilde\sigma^a,N] ,\nonumber\\
\{C_N,A_{a } \}&=&{\cal D}_a N ,\nonumber\\
\{C\;{}_{\hbox {${}_{{}_{{}_{\widetilde{}}}}$\kern-.5em{\small\it
N}}},\tilde\sigma^a{} \}&=&2\;{\cal D}_b \;(\;\hbox
{${}_{{}_{{}_{\widetilde{}}}}$\kern-.5em\it 
N}\tilde\sigma^{[a}\tilde\sigma^{b]} ) ,\\
\label{poisson}
\{C\;{}_{\hbox {${}_{{}_{{}_{\widetilde{}}}}$\kern-.5em {\small \it
N}}},A_{a}\}&=&\hbox {${}_{{}_{{}_{\widetilde{}}}}$\kern-.5em\it  N}\;[
\tilde\sigma^b, F_{ba}],\nonumber\\
\{C_{\vec N},\tilde\sigma^a{} \}&=&-{\cal L}_{\vec N}\tilde\sigma^a
,\nonumber\\
\{C_{\vec N},A_{a }\}&=&-{\cal L}_{\vec N} A_{a},\nonumber
\end{eqnarray}
and the constraints algebra becomes

\begin{eqnarray}
\label{algebra}
\{C_N,C_M\}&=&C_{[N,M]} ,\nonumber\\
\{C_{\vec N}, C_M \}&=&C_{{\cal L}_{\vec N} M},\nonumber\\
\{C_{\vec N},C_{\vec M}\}&=&C_{[\vec N,\vec M]},\nonumber\\ 
\{C_N, C\;{}_{\hbox {${}_{{}_{{}_{\widetilde{}}}}$\kern-.5em{\small\it
M}}} \}&=&0,\\
\{C_{\vec N}, C\;{}_{\hbox
{${}_{{}_{{}_{\widetilde{}}}}$\kern-.5em{\small\it M}}} \}&=&C_{{\cal
L}_{\vec N}\;\hbox {${}_{{}_{{}_{\widetilde{}}}}$\kern-.5em{\small\it 
M}}} \nonumber\\
\{C\;{}_{\hbox {${}_{{}_{{}_{\widetilde{}}}}$\kern-.5em{\small\it N}}}
,C\;{}_{\hbox {${}_{{}_{{}_{\widetilde{}}}}$\kern-.5em {\small \it M}}}
\}&=&
C_{\vec K}+C_{A_mK^m},\nonumber
\end{eqnarray}
where $K^a=2\sigma^2\;(\,\hbox
{${}_{{}_{{}_{\widetilde{}}}}$\kern-.5em \it M} {\cal D}^a\;\hbox
{${}_{{}_{{}_{\widetilde{}}}}$\kern-.5em \it N}-\hbox
{${}_{{}_{{}_{\widetilde{}}}}$\kern-.5em \it N} {\cal D}^a\;\hbox
{${}_{{}_{{}_{\widetilde{}}}}$\kern-.5em \it M})$.

Using the canonical variables $(\sigma^a,A_a)$ and the action, the
resulting \newline Hamiltonian writes:
\begin{eqnarray}
\label{eq:H}
H&=&\int_\Sigma \mbox{d}^3x \left(\;i\;\sqrt{2}\mbox{tr}( N^a
\tilde\sigma^bF_{ab} - N^a A_a{\cal D}_b\tilde\sigma^b)+\hbox
{${}_{{}_{{}_{\widetilde{}}}}$\kern-.5em\it
N}\;\;\mbox{tr}(\;\tilde\sigma^a\tilde\sigma^b
F_{ab})\right),\nonumber\\
&=&-C_{\vec N}+\frac{i}{\sqrt 2}C\;{}_{\hbox
{${}_{{}_{{}_{\widetilde{}}}}$\kern-.5em {\small \it N}}},
\end{eqnarray}
where the Lagrange multiplier ${}^4A_a t^a$ has been chosen as $A_a
N^a$.

Therefore, the Hamiltonian evolution of the dynamical variables, yields:
\begin{eqnarray}
\label{eq:evolutionS}
{\cal L}_t\tilde\sigma^b&=&\{H,\tilde\sigma^b\}\nonumber\\
&=&{\cal L}_{\vec N}\tilde\sigma^b-\frac{i}{\sqrt{2}}{\cal D}_a(\;\hbox
{${}_{{}_{{}_{\widetilde{}}}}$\kern-.5em\it
N}[\tilde\sigma^{a},\tilde\sigma^{b}])\nonumber\\
&=&-[A_a N^a,\tilde\sigma^b]+2{\cal D}_a(N^{[a}\tilde\sigma^{b]})+N^b
{\cal D}_a \tilde\sigma^{a}
-\frac{i}{\sqrt{2}}{\cal D}_a(\;\hbox
{${}_{{}_{{}_{\widetilde{}}}}$\kern-.5em\it
N}[\tilde\sigma^{a},\tilde\sigma^{b}]), \\
\label{eq:evolutionA}
{\cal L}_t A_b&=&\{H,A_b\}\nonumber\\
&=&{\cal L}_{\vec N}A_b+\frac{i}{\sqrt{2}}\;\;\hbox
{${}_{{}_{{}_{\widetilde{}}}}$\kern-.5em\it
N}[\tilde\sigma^a,F_{ab}]\nonumber\\
&=&{\cal D}_b(A_a N^a)+N^aF_{ab}+\frac{i}{\sqrt{2}}\;\;\hbox
{${}_{{}_{{}_{\widetilde{}}}}$\kern-.5em\it N}[\tilde\sigma^a,F_{ab}].
\end{eqnarray}
Note that the evolution obtained from the Lagrangian differs from the
evolution obtained from the Hamiltonian in a "constraint term".

%%%%%%%%%%%%%%%%%%%%%%%%%%%%%%%%%%%%%%%%%%%%%%%%%%%%%%%%%%%%%%%%%%%%%%%%%%%%%%%%%%%%%%%%%%%%%%%%%
\begin{center}
{\bf APPENDIX C: THE CONSTRAINTS EVOLUTION}\end{center}
\label{evolconstr}

To compute the evolution equations for the constraints with the extended 
flow let us first calculate the Hamiltonian evolution of the constraints
$(\tilde C_A{}^B, C, C_a)$. 
Using the Hamiltonian given by
(\ref{eq:H}), the constraint algebra (\ref{algebra}) and integrating
by parts, we obtain
\begin{eqnarray}
\label{eq:constraintevol}
\dot{\tilde C}&=&\{H,\tilde C\}={\cal L}_{\vec N}\tilde C, \nonumber\\
\dot{C}&=&\{H, C\}={\cal L}_{\vec N}C+i\sqrt{2}\;\sigma^2\;\hbox
{${}_{{}_{{}_{\widetilde{}}}}$\kern-.5em\it N}\;{\cal D}^a C_a
+i\;2\sqrt{2}\sigma^2 C^a{\cal D}_a \;\hbox
{${}_{{}_{{}_{\widetilde{}}}}$\kern-.5em\it N}.\\
\dot{C}_a&=&\{H, C_a\}={\cal L}_{\vec N} C_a-\frac{i}{\sqrt{2}}\;\hbox
{${}_{{}_{{}_{\widetilde{}}}}$\kern-.5em\it N}\;{\cal D}_a C-i\sqrt{2}
C {\cal D}_a \;\hbox {${}_{{}_{{}_{\widetilde{}}}}$\kern-.5em\it
  N}+\frac{i}{\sqrt{2}}\;\hbox
{${}_{{}_{{}_{\widetilde{}}}}$\kern-.5em\it
  N}\;\mbox{tr}([\tilde\sigma^b, F_{ba}]\tilde C).\nonumber
\end{eqnarray}

The most complicated calculation is the last one (the evolution of $C_a$), in order to do this,  we define the
functional
$$
\hat C_{\vec M}=-i\sqrt{2}\int_\Sigma \mbox{d}^3x
\;M^a\;\mbox{tr}(\tilde\sigma^bF_{ab}),
$$
then
\begin{equation}
\label{eq:conmut1}
\{H,\hat C_{\vec M}\}=-\{C_{\vec N}, \hat C_{\vec
M}\}+\frac{i}{\sqrt{2}}\{C_{\hbox
{${}_{{}_{{}_{\widetilde{}}}}$\kern-.5em {\small\it N}}}, \hat C_{\vec
M}\}.
\end{equation}
We calculate each one of the terms above as follows
\begin{eqnarray}
\label{eq:conmut2}
\{C_{\vec N},\hat C_{\vec M}\}&=&\frac{i}{\sqrt{2}}\int_\Sigma
\mbox{d}^3x\; \mbox{tr}\left(\frac{\delta {C_{\vec N}}}{\delta
A_a}\frac{\delta \hat C_{\vec M}}{\delta \sigma^a}-\frac{\delta {C_{\vec
N}}}{\delta \sigma^a}\frac{\delta \hat C_{\vec M}}{\delta
A_a}\right),\nonumber\\
&=&\{C_{\vec N}, C_{\vec M}\}+\{C_{\vec
N},C_{P}\}|_{P=M^aA_a}\nonumber \\ 
&&+i\sqrt{2}\int_\Sigma\mbox{d}^3x\;{\cal L}_{\vec
N}A_a\;M^a C,\nonumber \\
&=&C_{[\vec N, \vec M]}+C_{{\cal L}_{\vec N} (M^aA_a})+,C_{M^a{\cal
L}_{\vec N} A_a}\nonumber\\
&=&C_{[\vec N, \vec M]}+C_{[\vec N,\vec M]^aA_a},\nonumber\\
&=&\hat C_{[\vec N,\vec M]},
\end{eqnarray}
and
\begin{eqnarray}
\label{eq:conmut3}
\{C{}_{\hbox {${}_{{}_{{}_{\widetilde{}}}}$\kern-.5em {\small\it
N}}},\hat C_{\vec M}\}&=&\{C{}_{\hbox
{${}_{{}_{{}_{\widetilde{}}}}$\kern-.5em {\small\it N}}},C_{\vec
M}\}+\{C{}_{\hbox {${}_{{}_{{}_{\widetilde{}}}}$\kern-.5em {\small\it
N}}}, C_{P}\}|_{P=M^aA_a}\nonumber \\
&&-i\sqrt{2}\int_\Sigma\mbox{d}^3x\;\hbox
{${}_{{}_{{}_{\widetilde{}}}}$\kern-.5em \it
N}\;\mbox{tr}([\tilde\sigma^b,F_{ab}]M^a \tilde C),\nonumber\\
&=&-C{}_{{\cal L}_{\vec M}\;\hbox
{${}_{{}_{{}_{\widetilde{}}}}$\kern-.5em {\small\it N}}}+C{}_{\hbox
{${}_{{}_{{}_{\widetilde{}}}}$\kern-.5em {\small\it
N}}\;[\tilde\sigma^b,F_{ab}]M^a}.
\end{eqnarray}
Inserting the Poisson brackets (\ref{eq:conmut2}) and
(\ref{eq:conmut3}) in (\ref{eq:conmut1}), using that the Lie
derivative of a function with weight minus one is ${\cal L}_{\vec
  M}\;\hbox {${}_{{}_{{}_{\widetilde{}}}}$\kern-.5em \it
  N}=M^a\partial_a \;\hbox {${}_{{}_{{}_{\widetilde{}}}}$\kern-.5em
  \it N}-\;\hbox {${}_{{}_{{}_{\widetilde{}}}}$\kern-.5em \it
  N}\;\partial_a M^a$ and integrating by parts we get
$$
\{H, C_a\}={\cal L}_{\vec N} C_a-\frac{i}{\sqrt{2}}\;\hbox
{${}_{{}_{{}_{\widetilde{}}}}$\kern-.5em\it N}\;{\cal D}_a C-i\sqrt{2}
C {\cal D}_a \;\hbox {${}_{{}_{{}_{\widetilde{}}}}$\kern-.5em\it
  N}+\frac{i}{\sqrt{2}}\;\hbox
{${}_{{}_{{}_{\widetilde{}}}}$\kern-.5em\it
  N}\;\mbox{tr}([\tilde\sigma^b, F_{ba}]\tilde C).
$$
Since we have changed the evolution of the dynamical variables
outside the constraint sub-manifold, the calculus of the time
derivative of the constraint equations must be done using equations
(\ref{eq:evolution1S}) and (\ref{eq:evolution1A}). If $f:\Gamma
\to {\bf C}$ then
$$
\dot f(\tilde \sigma,A)=\int_{\Sigma} d^3x
\quad\mbox{tr}\left(\frac{\delta f}{\delta \tilde\sigma^a}
\dot{\tilde \sigma}^a+ \frac{\delta f}{\delta A_a} \dot{A}_a \right);
$$
and since the perturbation we have made on the equations
(\ref{eq:evolution1S}) and (\ref{eq:evolution1A}) are linear in the
constraints, we obtain

\begin{eqnarray}
\label{eq:derivtemp}
&&\dot f(\tilde \sigma,A)=\{H,f\}+\int_{\Sigma} d^3x\nonumber\\
&& \times\mbox{tr}\left(\frac{\delta
  f}{\delta \tilde\sigma^b} \left(\frac{i}{\sqrt{2}}\;\;\hbox
  {${}_{{}_{{}_{\widetilde{}}}}$\kern-.5em\it N} [\tilde
  C,\tilde\sigma^b]\right)- \frac{\delta f}{\delta A_b}
\left(\frac{i}{\sigma^2\sqrt{2}}\;\;\hbox
  {${}_{{}_{{}_{\widetilde{}}}}$\kern-.5em\it N} \tilde\sigma_b
  C+\frac{i}{\sigma^4}\;\;\hbox
  {${}_{{}_{{}_{\widetilde{}}}}$\kern-.5em\it N}
  \tilde\epsilon_b{}^{dc}\tilde\sigma_c C_d\right)\right).
\end{eqnarray}
Then, using (\ref{eq:constraintevol}) in (\ref{eq:derivtemp}), we have
the  constraints evolution 
\begin{eqnarray}
\label{eq:constraintevol1}
\dot{\tilde C}&=&N^a\partial_a \tilde C -\frac{i}{\sqrt{2}} \;\hbox
{${}_{{}_{{}_{\widetilde{}}}}$\kern-.5em\it N}\;[\tilde\sigma^a,{\cal
D}_a \tilde C]-\frac{i}{\sqrt{2}} \;{\cal D}_a \;\hbox
{${}_{{}_{{}_{\widetilde{}}}}$\kern-.5em\it N}\;[\tilde\sigma^a, \tilde
C]-i2\sqrt{2}\;\hbox {${}_{{}_{{}_{\widetilde{}}}}$\kern-.5em\it N}\;C_d
\tilde\sigma^d,\nonumber\\
\dot{ C}&=&N^a D_a  C- i\sqrt{2} \;\hbox
{${}_{{}_{{}_{\widetilde{}}}}$\kern-.5em\it N} \sigma^2 D^a
C_a-i\sqrt{2}C^d\;\hbox {${}_{{}_{{}_{\widetilde{}}}}$\kern-.5em\it
N}\;\makebox{tr}\left(\tilde C
\tilde\sigma_d\right)\nonumber\\
&&+\frac{i}{\sqrt{2}}\;\hbox
{${}_{{}_{{}_{\widetilde{}}}}$\kern-.5em\it
N}\;\makebox{tr}\left([\tilde C,
\tilde\sigma^a][\tilde\sigma^b,F_{ab}]\right)+\frac{i}{\sigma^4}\;\hbox {${}_{{}_{{}_{\widetilde{}}}}$\kern-.5em\it
N}\;C \tilde \epsilon^{abe}\makebox{tr}\left(\left({\cal D}_b
\tilde\sigma_a\right)\tilde\sigma_e\right),\\
\dot{ C_a}&=&N^bD_b  C_a+\frac{i}{\sigma^2}\;\hbox
{${}_{{}_{{}_{\widetilde{}}}}$\kern-.5em\it N}\;\tilde
\epsilon_a{}^{bd}D_b C_d+\frac{i}{\sqrt{2}}\;\hbox
{${}_{{}_{{}_{\widetilde{}}}}$\kern-.5em\it N}\;D_a C+C_b D_a
N^b\nonumber\\
&&-\frac{i}{\sigma^2}C_d D_b \;\hbox
{${}_{{}_{{}_{\widetilde{}}}}$\kern-.5em\it N}\;\tilde
\epsilon_a{}^{db}-\frac{2i}{\sigma^4}\;\hbox
{${}_{{}_{{}_{\widetilde{}}}}$\kern-.5em\it N}\;\makebox{tr}\left(
\left({\cal D}_{[a} \tilde\sigma_{|c|}\right)\tilde\sigma^b\right)\tilde
\epsilon_{b]}{}^{dc} C_d.\nonumber
\end{eqnarray}

%%%%%%%%%%%%%%%%%%%%%%%%%%%%%%%%%%%%%%%%%%%%%%%%%%%%%%%%%%%%%%%%%%%%%%%%%%%%%%%%%%%%%%%%%%%%%%%%%%%%%
%%%%%%%%%%%%%%%%%%%%%%%%%%%%%%%%%%%%%%%%%%%%%%%%%%%%%%%%%%%%%%%%%%%%%%%%%%%
\begin{center}
{\bf APPENDIX D: THE REALITY CONDITIONS}
\end{center}
\label{real}

Consider the variables $(\tilde\sigma^a,A_a)$. We want to prove that if
they define a real metric 
on an initial surface, then  this metric remains real under the
Hamiltonian evolution. In order to do this, we shall calculate the
evolution of the variables 
$q_{ab}= -$tr $\left (\sigma_a\sigma_b\right)$ and 
$\pi_{ab}=-$tr $\left (\pi_a\sigma_b\right)$, where $\pi_a$ is defined
by
$$
{\cal D}_a\lambda_A=D_a\lambda_A+\frac{i}{\sqrt 2}
\pi_{aA}{}^B\lambda_B.
$$
We write the  evolution of $\tilde \sigma_a$ in terms of $\pi_{ab}$ as
follows

\begin{eqnarray}
\label{eq:evolutionS10}
{\cal L}_t\tilde\sigma^b
&=&{\cal L}_{\vec N}\tilde\sigma^b-\frac{i}{\sqrt{2}}{\cal
D}_a\left(\;\hbox {${}_{{}_{{}_{\widetilde{}}}}$\kern-.5em\it
N}[\tilde\sigma^{a},\tilde\sigma^{b}]\right)+\frac{i}{\sqrt{2}}\;\hbox
{${}_{{}_{{}_{\widetilde{}}}}$\kern-.5em\it N}[{\cal
D}_a\tilde\sigma^{a},\tilde\sigma^{b}]\nonumber\\
&=&{\cal L}_{\vec N}\tilde\sigma^b-i\;{\cal D}_c\;\hbox
{${}_{{}_{{}_{\widetilde{}}}}$\kern-.5em\it
N}\;\sigma\;\epsilon^{cbe}\;\tilde\sigma_e+\sigma\;\hbox
{${}_{{}_{{}_{\widetilde{}}}}$\kern-.5em\it
N}\left(\pi^b{}_e\tilde\sigma^e-\pi \tilde\sigma^b\right),
\end{eqnarray}
then we calculate
\begin{eqnarray}
\label{eq:evolutionqabt}
{\cal L}_t\tilde{q}^{ab}&=&-2\mbox{tr}\left({\cal
L}_t\tilde\sigma^{(a}\tilde\sigma^{b)}\right)\nonumber\\
&=&{\cal L}_{\vec N}\tilde{q}^{ab}+2\;\hbox
{${}_{{}_{{}_{\widetilde{}}}}$\kern-.5em\it
N}\sigma^3\left(\pi^{(ab)}-\pi q^{ab}\right).
\end{eqnarray}
The evolution of $q_{ab}$ follows from the evolution of 
$\sigma^2\equiv \mbox{det}\left(q_{ab}\right)$, thus we calculate
\begin{equation}
\label{eq:evolutiondet}
{\cal L}_t\sigma^2=2\sigma^2 D_a N^a-2\sigma^3\pi\;\hbox
{${}_{{}_{{}_{\widetilde{}}}}$\kern-.5em\it N}.
\end{equation}
Finally, we get
\begin{equation}
\label{eq:evolutionqab}
{\cal L}_t q_{ab}
={\cal L}_{\vec N} q_{ab}-2N \pi_{(ab)}.
\end{equation}
Thus in order to ensure the reality of $q_{ab}$ we need to know the
reality of $\pi_{ab}$. 
The evolution of $\pi_{ab}$ is given by
$$
{\cal L}_t\pi_{ab}=-\mbox{tr}\left(\pi_a{\cal
L}_t\sigma_b\right)+i\sqrt{2}\mbox{tr}\left(\sigma_b{\cal L}_t
A_a\right)-i\sqrt{2}\mbox{tr}\left(\sigma_b{\cal L}_t \Gamma_a\right).
$$
Hence in order to calculate it,   we need to rewrite the evolution of $A_b$ and  $\sigma_a$ in terms
of $\pi_{ab}$. Using 
$$
F_{ab}=R_{ab}-\pi_{[a}\pi_{b]}+i\sqrt{2}D_{[a}\pi_{b]}
$$
where $R_{ab}$ is the spinorial curvature, and  redefining $C=\mbox{tr}\left(\sigma^a\sigma^b F_{ab}\right)$ and
$C_b=\mbox{tr}\left(\sigma^aF_{ab}\right)$, we compute

\begin{eqnarray}
\label{eq:evolutionS11}
{\cal L}_t\sigma_b
&=&{\cal L}_{\vec N}\sigma_b-i\; D_c
N\;\sigma\;\epsilon^c{}_{be}\;\sigma^e-N \pi_{eb} \sigma^e\\
\label{eq:evolutionA10}
{\cal L}_t A_b&=&{\cal L}_{\vec N}A_b+\frac{i}{\sqrt{2}}
N[\sigma^a,F_{ab}] -\frac{i}{\sqrt{2}} N \sigma_b C-i N
\epsilon_b{}^{dc}\sigma_c C_d\nonumber\\
&=&{\cal L}_{\vec N}A_b+\frac{i}{\sqrt{2}} N[\sigma^a,R_{ab}]
\nonumber\\
&&-\frac{i}{\sqrt{2}}
N\left(\pi_{an}\pi_b{}^a-\pi\pi_{bn}\right)\sigma^n-N[\sigma^a,D_{[a}\pi_{b]}]\nonumber\\
&&-\frac{i}{\sqrt{2}}
N \sigma_b C-i N \epsilon_b{}^{dc}\sigma_c C_d.
\end{eqnarray}
Hence  
\begin{eqnarray}
\label{eq:evolutionpiab}
{\cal L}_t\pi_{ab} &=&{\cal L}_{\vec
N}\pi_{ab}+N\pi\pi_{ab}-2N\pi_{eb}\pi_a{}^e-N R_{ab}-q_{ab} N
C+N\sqrt{2}\epsilon_{ab}{}^d C_d\nonumber\\
&&-i2N\epsilon_b{}^{de}\mbox{tr}\left(D_{[d}\pi_{a]}\sigma_e\right)+iD_cN\epsilon^{ce}{}_b\pi_{ae}\nonumber\\
&&+i\sqrt
2\mbox{tr}\left(\left({\cal L}_{\vec N}\Gamma_a-{\cal
L}_{t}\Gamma_a\right)\sigma_b\right).\end{eqnarray}
Using the fact that $D_a\sigma_b=0$ and  the formula \ref{eq:gamma} for  $\Gamma_a$,     the  last three  terms can
be set as
\begin{eqnarray*}
&&-i2N\epsilon_b{}^{de}\mbox{tr}\left(D_{[d}\pi_{a]}\sigma_e\right)+iD_cN\epsilon^{ce}{}_b\pi_{ae}+i\sqrt
2\mbox{tr}\left(\left({\cal L}_{\vec N}\Gamma_a-{\cal
L}_{t}\Gamma_a\right)\sigma_b\right)\\
&=&-D_aD_bN-iN\epsilon_b{}^{dm} D_a\pi_{dm}-\frac{i}{2}q_{ab}\epsilon^{cmd}\left(D_cN\pi_{md}+ND_c\pi_{md}\right),
\end{eqnarray*}
yielding
\begin{eqnarray}
\label{eq:realconda}
{\cal L}_t \pi_{ab}&=&{\cal L}_{\vec
N}\pi_{ab}+N\pi\pi_{ab}-2N\pi_a{}^e\pi_{eb}-N
R_{ab}-D_aD_bN\nonumber\\
&&-\frac{N}{2}q_{ab}\left(R+\pi^2-\pi_{dc}\pi^{cd}\right)-iN\epsilon_{ab}{}^d D^c\left(\pi_{dc}-\pi
q_{dc}\right)+\nonumber\\
&&2N\left(\pi^d{}_a\pi_{[db]}-\pi\pi_{[ab]}+\pi_{[ad]}\pi^d{}_b\right)\nonumber\\
&&-iN\epsilon_b{}^{dm} D_a\pi_{[dm]}-\frac{i}{2}q_{ab}\epsilon^{cmd}\left(D_cN\pi_{[md]}-ND_c\pi_{[md]}\right).
\end{eqnarray}

%%%%%%%%%%%%%%%%%%%%%%%%%%%%%%%%%%%%%%%%%%%%%%%%%%%%%%%%%%%%%%%%%%%%%%%%%%%%%


\begin{thebibliography}{99}



\bibitem{cho}Y. Bruha {\it Theoreme d'existence pour certain systemes d'equations aux derive\'es partielles nonlinaires}, Acta Math., {\bf 88}, 141-225, (1952).

\bibitem{cho-rugg}Y. Choquet-Bruhat and T. Ruggeri.{\it Hyperbolicity of the 3+1 System of Einstein Equations}, Comm. Math. Phys., {\bf 89}, 269-275, (1983).

\bibitem{helmut} H. Friedrich, {\it The asymptotic characteristic initial value problem for Einstein's vacuum field
equations as an initial value problem for a first-order quasilinear
symmetric hyperbolic
     system}, Proc. Roy. Soc. A, 378, 401-421, (1981). {\it On the hyperbolicity of Einstein's and other gauge
field equations}, Commun.
     Math. Phys., {\bf 100}, 525-543, (1985). {\it Hyperbolic reductions for Einstein's equations},  Class. Quantum Grav., {\bf 13}, 1451-1469, (1996).

\bibitem{fri-reu}S. Fritelli and O. Reula, Comm. Math. Phys.,{\it On the newtonian limit of general relativity},   {\bf 166}, 2, (1994). {\it First-order symmetric-hyperbolic
Einstein equations with
     arbitrary fixed gauge}, Phys. Rev. Lett., 76, 4667-4670, (1996).


\bibitem{bonna-masso} C. Bona and J. Mass\'o, Phys. Rev. Letter, {\bf 68}. 1097, (1992).

\bibitem{cho-york}A. Abrahams, A. Anderson, Y.Choquet-Bruhat and J. York,{\it Geometrical hyperbolic systems for general
relativity and gauge theories}, Class Quantum Grav., {\bf 14}, A9-A22,
(1997).

\bibitem{putten} 0 van Putten, M.H.P.M., and Eardley, D.M., {\it Nonlinear wave equations
for relativity}, Phys.
     Rev. D, 53, 3056-3063, (1996). 
 
\bibitem{livrev}O. Reula,{\it Hyperbolic Methods for Einstein's Equations}, Living Reviews in Relativity,
http:// www.livingreviews.org/Articles/Volume1/1998-3reula/, 1998.

\bibitem{Iriondo}M. Iriondo, E. Leguizamon, O. Reula{\it Einstein's equations in Ashtekar's variables constitute a symmetric hyperbolic system}, Physical Review Letters,{\bf 79},  4132, 1997.
 

\bibitem{ash}A. Ashtekar,{\it New Perspectives in Canonical Gravity}, edited by Bibliopolis, Naples, 1988.

\bibitem{ash-rom} A. Ashtekar,  R. Romano,  S. Tate {\it New variables for gravity:Inclusion of matter}, Physical Review D, {\bf 40}, 8,2572, 1989.

\bibitem{immirzi}G. Immirzi {\it The reality conditions for the new canonical
variables of general relativity}, Class. Quan. Grav. {\bf 10}, 2347-2352, 1993.

\bibitem{simo}S. Frittelli {\it Note on the propagation of the constraints in
standard 3+1 general relativity.} Phys. Rev. D, {\bf 55}, 5992-5996, 1997.



\end{thebibliography}
\end{document}